\documentclass[apj]{emulateapj}

\usepackage{natbib}
\usepackage{graphicx}
\bibpunct{(}{)}{;}{a}{}{,}

\begin{document}

   \title{Automated unsupervised classification of the Sloan Digital Sky Survey stellar spectra
using k-means  clustering
}
   \author{
          	J.~S\'anchez~Almeida\altaffilmark{1,2}
                and
         	C.~Allende~Prieto\altaffilmark{1,2}
          }
\altaffiltext{1}{Instituto de Astrof\'\i sica de Canarias, E-38205 La Laguna,
Tenerife, Spain}
\altaffiltext{2}{Departamento de Astrof\'\i sica, Universidad de La Laguna,
Tenerife, Spain}
\email{jos@iac.es, callende@iac.es}
\begin{abstract}
Large spectroscopic surveys  require automated methods of analysis.   
This paper explores the use of  k-means 
clustering  as a tool for automated 
unsupervised classification of massive stellar spectral catalogs.
The classification criteria are defined by the data and the algorithm, 
with no prior physical  framework. We work with a representative set of
stellar spectra associated with the SDSS SEGUE and SEGUE-2 programs, which 
consists of 173,390 spectra from  3800 to 9200\,\AA\  sampled on 
3849 wavelengths. We classify the original spectra as well as 
the spectra with the continuum removed. The second set 
only contains spectral lines, and it is less dependent on
uncertainties of the flux calibration.
The classification of the spectra with continuum renders
16 major classes. Roughly speaking,  
stars are split according to their colors, with enough 
finesse to distinguish dwarfs from giants of the same 
effective temperature, but with difficulties to separate 
stars with different metallicities.  
There are classes corresponding to particular MK types,
intrinsically blue stars dust-reddened,
stellar systems, and also classes collecting faulty spectra. 
Overall, there is no one-to-one correspondence
between the classes we derive and the MK types.
The classification of spectra without continuum renders 13 classes,
the color separation is not so sharp, but it distinguishes stars
of the same effective temperature and different metallicities.
Some classes thus obtained present a fairly small range of physical 
parameters (200~K in effective temperature,
0.25 dex in surface gravity, and 0.35 dex in metallicity),  
so that the classification can be used to estimate
the main physical parameters of some stars 
at a minimum computational cost.
We also analyze the outliers of the classification. 
Most of them turn out to be failures of the reduction pipeline, but there
are also high redshift QSOs, multiple stellar systems, dust-reddened stars,  
galaxies, and, finally, odd spectra whose nature we have not decipher. 
The template spectra representative of the classes are publicly available
({\tt ftp://stars:kmeans@ftp.iac.es}). 
\end{abstract}




%
%
\section{Introduction}\label{intro}

Stellar spectra contain a wealth of information on the photospheres of
stars, including their chemical makeup. In spite of decades of measuring and studying stellar spectra, 
we have a very limited knowledge of how many stars with a given chemical composition exist
in the Galaxy. Large spectroscopic surveys such as RAVE \citep[][]{2006AJ....132.1645S} or 
SDSS \citep[e.g.,][]{2002AJ....123..485S}
have increased by several orders of magnitude the number of stars with measured spectra, 
yet these surveys are limited in scope,  reaching only a particular cross-section of the stellar population 
of the Milky Way,  and biased, in that they target stars that have been preselected based on their 
color, distance or brightness.  This situation will soon change as new large projects such as 
Gaia \citep[e.g.,][]{2005ESASP.576.....T}, or  HETDEX \citep{2008ASPC..399..115H}, with 
well controlled samples,  appear in the scene.
The advent of these new data sets poses an obvious problem.
As the data flow continuously increases, performing systematic studies becomes 
more of an issue, and new efficient 
means of analyzing  stellar spectra are needed.

The study of stellar spectra to quantify the physical properties 
of stars requires model atmospheres, radiative transfer calculations, atomic and molecular data, 
and involves a large number of approximations.
The model ingredients and our recipes to apply them to interpret observations are in 
constant evolution, and so are the values for the inferred quantities. Avoiding such changes is 
one of the motivations behind  spectroscopic classification in general, and in particular of the MK 
system of \citet{1943QB881.M6.......} and \citet{1973ARA&A..11...29M} 
which is still in use today enhanced with extensions. 
It assigns spectral classes in a way that is purely empirical and repeatable,  
providing basic information as a  preliminary step for further,  
more detailed, analysis.

This kind of typing system based on a series of predefined criteria 
is feasible as long as the criteria are set and applied by humans. Such systems
resemble the taxonomical classification of animal species \citep[see][]{2006A&A...455..845F}
and,  by definition, 
are not optimal since they rely on subjective judgments. 
If such systems have to be updated for application
to very large number of spectra, the classification has to be  
easily implemented and performed with computers, in a fast 
and homogeneous fashion. 
Moreover, \citet{san05} maintained 
that physics must not be used to drive a classification; 
otherwise the arguments become circular when using the
classification to support physics. Thus, one is inclined 
to consider unsupervised classification 
systems, which are themselves defined by algorithms and data.
The potentials of one of such methods, k-means, 
is explored in this paper. However, it must be stressed 
that {\em unsupervised} does not mean {\em absolute} or 
model {\em independent}. The classification criteria are 
implicitly set by the algorithm,  and  the resulting classes 
depend on the specific dataset under analysis.  

In data mining parlance, the spectrum of a star
is a point in the high-dimensional space where each
coordinate corresponds to the intensity at a particular 
wavelength.
Given a comprehensive set of stellar spectra, 
classifying  consists of  identifying clusters
in this high-dimensional space.
The problem of finding structures in a multidimensional data set goes also by the 
name of cluster analysis \citep[see e.g.,][]{eve95,bis06}. One of the
most widely used algorithms is k-means clustering \citep{mac67}, 
and it fulfills the requirements  put forward above.
Moreover, k-means is simple to code and robust, even when exploring clustering 
in a high-dimensional space. 
Previous works have shown successful applications 
of the method to the classification of spectra in
various astrophysical contexts, e.g.,
stars \citep[][]{1996A&A...311..145B,2012arXiv1209.0495S}, 
solar polarization spectra \citep{2000ApJ...532.1215S,2011A&A...530A..14V},
X-ray spectra \citep{2007ApJ...659..585H}, 
spectra from asteroids \citep{2008AIPC.1082..165G}, 
and galaxy spectra \citep{2009ApJ...698.1497S,2010ApJ...714..487S,2011ApJ...743...77M}.

We now explore its application to medium-resolution stellar 
spectra from the Sloan Digital Sky Survey.
SDSS currently provides the largest available homogenous database of stellar spectra.
The
original SDSS survey \citep{2002AJ....123..485S,2009ApJS..182..543A} 
together with 
SEGUE and SEGUE-2 contain  
somewhere over half a million stellar spectra
\citep{2009AJ....137.4377Y}. 
The Baryon Oscillations Spectroscopic Survey 
\citep[BOSS;][]{2011AJ....142...72E}, 
part of SDSS-III, uses similar but upgraded spectrographs, and 
in the first two years of operation has already obtained over 100,000 additional spectra of stars. 
Even though these stars do not provide a fair sample of the Milky Way stellar population, 
this rich data set is a good place to explore the application of clustering algorithms 
for classifying stars.
Actually, the set has  already been used  for this
purpose.  \citet{2010AJ....139.1261M} apply principal component 
analysis (PCA) to spectra having narrow color bins, so as to separate stars of the same effective 
temperature according to their gravity and metallicity. In this case the spectra are not 
classified directly by the automated procedure, but the color cuts
introduce human supervision into the classification.
\citet{2011AJ....142..203D} apply local linear embedding (LLE), which
is a type of PCA decomposition that preserves the 
nonlinear structure within high-dimensional data sets. The stellar spectra
are found to form a 1D family when projected into the first three eigenvectors.
Finally, \citet{2010ApJ...719.1759A} also use SDSS data to show  that stellar spectra 
are highly compressible, so that a small number of parameters suffice
to reproduce the bulk of the observed spectra.

In this paper we focus on the k-means  classification of stars observed as part of the SEGUE 
and SEGUE-2 surveys \citep{2009AJ....137.4377Y}.  
Section~\ref{data}  presents the selection of spectra used in the analysis, 
and Sect.~\ref{algorithm}  describes the basics behind k-means. 
Section~\ref{classification}  applies the algorithm to SDSS 
spectra, first considering continuum (Sect.~\ref{class_with}), and then
without continuum (Sect.~\ref{class_out}).
The classification allows us to identify outliers, often  
rare objects that do not show up unless the catalogs are large 
enough and which turn out to be extremely revealing 
\citep[e.g.,][]{2012ApJS..200...14M}.
The properties of these outliers are analyzed in  Sect.~\ref{outliers}.
Section~\ref{conclusions} explains the main results and outlines 
additional uses of the classification.

%
%
\section{Data set}\label{data}

The spectra come from SEGUE and SEGUE-2 \citep{2009AJ....137.4377Y}. These programs obtained 
stellar spectra using the SDSS 2.5-m telescope  and 
 the SDSS double spectrograph \citep{2006AJ....131.2332G,2012arXiv1208.2233S} between 
2004 and 2009. 
The spectra contain 3849 wavelengths 
covering the range 3800-9200\,\AA\ at a resolving power $R\simeq 1800$.
They  were  downloaded from the SDSS Data Release~8
\citep[DR8; ][]{2011ApJS..193...29A}.
SEGUE observed numerous types of targets, from very hot WD to very cool M and L~types, 
each chosen based on color criteria, and in some cases additional information such as proper motion. 
The survey observations sample the Galaxy at mid and high galactic latitudes, 
covering very sparsely 3/4 of the sky, with only 3 plates 
(less than 0.5\% of the spectra) at $|b|<10^\circ$.

The sample of DR8 spectra associated with SEGUE and SEGUE-2 programs
include 355,840 spectra in 525 plug-plates. Each plate, as for SDSS, admits 640 fibers. 
We selected stars with a median signal-to-noise ratio ${\rm S/N} >10$, a radial velocity module smaller 
than 600~km\,s$^{-1}$ (redshift $<0.002$), and which 
were not labeled as one of the
 following classes of objects: GALAXY, NA, QA, ROSAT\_D, QSO, SER, and CATY\_VAR.

We processed the spectra to eliminate by interpolation the [OIII] nightglow lines 
at 5577 and 6300\,\AA , and corrected the Doppler shifts associated with the radial velocities. 
The original spectra have units of flux per unit wavelength 
(i.e., erg\,cm$^{-2}$\,s$^{-1}$\,\AA$^{-1}$).  
We normalized them dividing
by the median value of their fluxes in the 
spectral band between 5000 and 6000~\AA . This step preserves the shape 
of the spectral energy distribution, while places all the spectra on the same 
scale regardless of the intrinsic luminosity of the stars, their distance, and the
amount of interstellar absorption. Missing sections of
spectra were patched by interpolation, since any regions with extreme (wrong) values can
damage the classification algorithm.
A sample 173,390 stellar spectra passed all the selection criteria, and 
we refer to them 
as {\em the reference set}.
Interstellar extinction is not corrected for, however, we analyze
a  version of the same data with a running mean subtracted from the spectral 
energy distribution, leaving only absorption features. This procedure is intended to 
partly remove the reddening produced by interstellar extinction, thus  minimizing its
potential impact on the results.

Two additional sets of spectra are mentioned in the paper. They were  
used only for testing in early stages of the
work, but they are explicitly mentioned here  because
the sanity checks performed with them provide confidence on the 
technique.
As far as we can tell, they are equivalent to the reference set 
in a statistical sense.
For lack of imagination, we refer to them as 
{\em 1st auxiliary set} and  {\em 2nd auxiliary set}.
Auxiliary set~1 comprises  63,611 stellar spectra from 3800 to 8000\,\AA , drawn
from SDSS/DR6 and then purged to retain only those with best S/N. They are uniformly 
resampled
in log-wavelength as to have 1617 wavelengths.  This set is particularly
rich in main sequence F-type stars, since they were used as spectrophotometric 
calibrators in the original SDSS survey (only 162 distinct SEGUE plates were included in DR6). 
Auxiliary set~2 comes from DR8, but the noise thresholding and other
selection criteria differ from the reference set, and leaves only 121,272 targets.

\begin{figure*}
\includegraphics[width=0.9\textwidth]{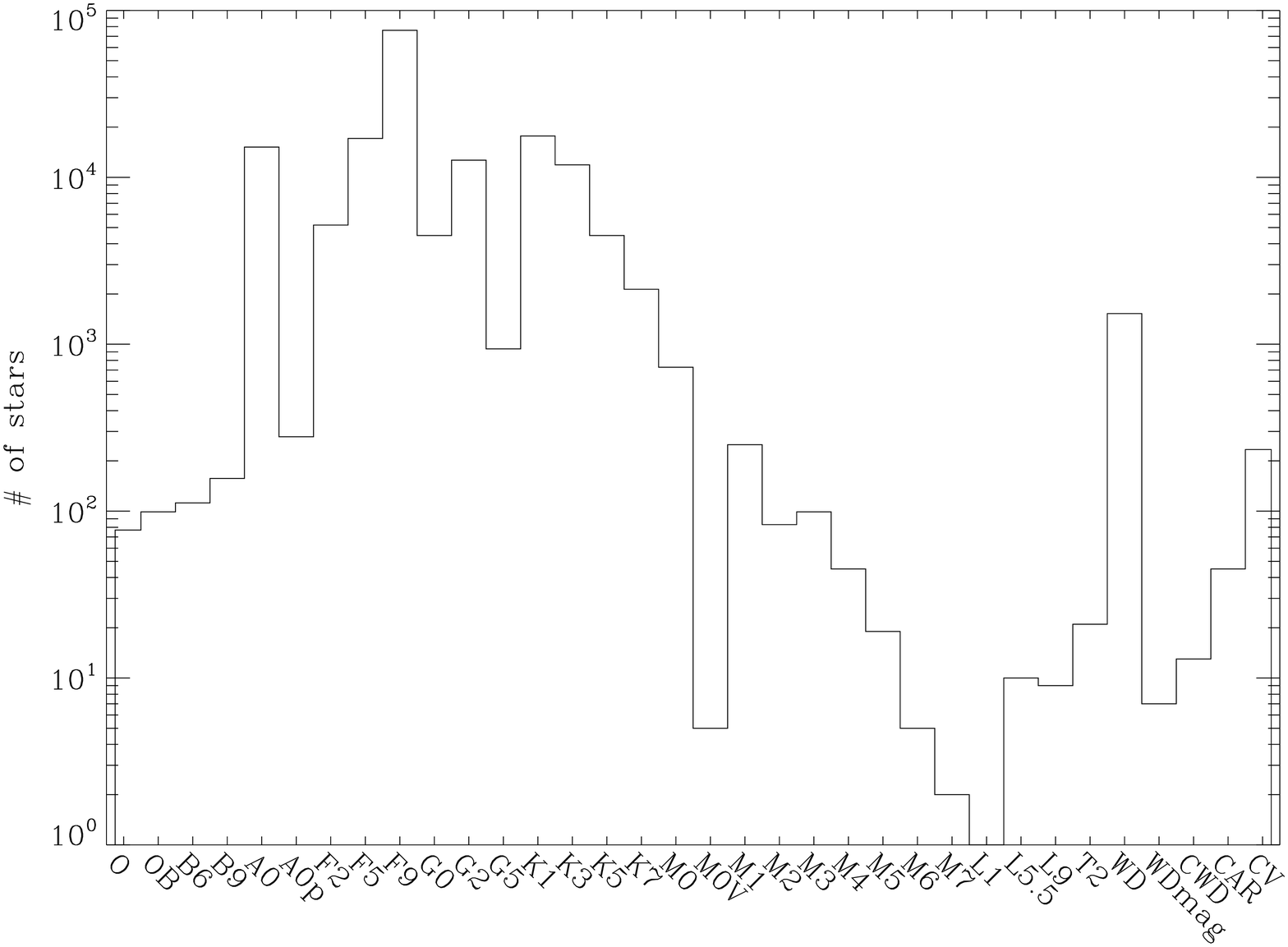}
\caption{
Distribution of MK spectral types in the reference dataset.
We use ELODIE-based types as provided by the SSPP.
}
\label{original_classes2}
\end{figure*}

Effective temperatures $T_{\rm eff}$, surface gravities ($\log {\rm g}$) and metallicities ([Fe/H]) computed
with  the SEGUE Stellar Parameter Pipeline 
\citep[hereafter SSPP;][]{
   2008AJ....136.2022L,
   2008AJ....136.2050L,
   2008AJ....136.2070A,
   2011AJ....141...90L,
   2011AJ....141...89S}
are used here to characterize the physical properties of the stars.
Each one of these physical parameters  results from the robust 
average of various independent  estimates,  discarding those which do not seem to 
be consistent \citep[for details, see][]{2008AJ....136.2022L}. 
The MK types of the targets mentioned in the paper 
are also from the SSPP. Most of our discussion 
is based on the so-called ELODIE MK types, derived 
by best-fitting templates from the ELODIE library 
of synthetic spectra \citep{2001A&A...369.1048P}.
The distribution of MK types of the reference
set is shown in Fig.~\ref{original_classes2}.
These types encompass a broad range of stellar
properties from O to L and include WDs, however, they 
face some difficulties when interpreting intermediate spectral types 
\citep[e.g., the distribution of MK types in Fig.~\ref{original_classes2} peaks at 
type F whereas most of the SEGUE stars have been selected to be of G type; see][]{2009AJ....137.4377Y}.  
The SSPP also provides a second set  of MK types only for cool stars, the Hammer MK types,
which were inferred using the spectral typing software developed and described by 
\citet{2007AJ....134.2398C}. They cope much better with types G and F, and
we apply them when appropriate.

\section{Classification algorithm}\label{algorithm}

We use {\em k-means} to carry out the classification, which is a robust tool
commonly used in data mining and artificial intelligence \citep[e.g.,][]{eve95,bra98}.
The actual realization of the algorithm employed in our analysis is described
in detail by \citet[][\S~2]{2010ApJ...714..487S}, and we refer to that work for details. 
However, for comprehensiveness, this section sketches the operation of 
the algorithm, with its pros and cons.  

As for most classification algorithms,  the stellar spectra are  vectors in 
a high dimensional linear space, with as many dimensions as
the number of wavelengths. Therefore the spectra to be classified are a set 
of points in this space. The points (i.e., the spectra) are assumed  to be 
clustered around  a number of centers.  Classifying consists of 
(a) finding the number of clusters and their centers, (b) 
assigning each spectrum to one of these centers, and
(c) estimating the probability that the choice is correct. 
The third step should be regarded as a sanity check that allows us to 
quantify the goodness of the classification for each particular spectrum. 
In the standard formulation, k-means starts by selecting at 
random from the full data set a number $k$ of  spectra. These spectra are 
assumed to be the center of a cluster. Then each spectrum of the data  set is assigned 
to the cluster center 
that is closest in a least squares sense\footnote{
This means using the Euclidean metric to assign distances between 
points in the high dimensional classification space.
Actually, we use the plain Euclidean distance,
where all the wavelengths are equally weighted 
\citep[][Eq.~(2)]{2010ApJ...714..487S}.
Observational errors are not included in the 
metric for simplicity. 
}.
Once all spectra have been assigned to one of the classes, the cluster centers are recomputed 
as the average of the spectra in the cluster. The procedure is iterated with the new 
cluster centers,  quitting when most spectra are not re-assigned in two 
successive steps  (99\,\% of them in our realization). 
The number of clusters is arbitrarily chosen but, the results are insensitive to 
such selection since only a few clusters possess a significant number of
members. Thus the algorithm provides the number of 
clusters, their corresponding cluster  centers, as well as the 
classification  of all the original spectra now 
assigned to one of the clusters. This information completes steps (a) and (b) of 
the classification procedure. 
In order to estimate the probability that the assignation is correct (step c), 
we compute for each cluster the distribution of the distances to the cluster center 
considering all spectra assigned to the cluster. 
We then assume that this distribution 
describes  the probability that any star with a given distance 
from the cluster center belongs to the class. Specifically,
the probability that a given star belongs to a cluster is estimated as the fraction 
of stars in the cluster with distances equal to or larger than the distance 
of the star.  
It is a sensible assumption;  it
gives high probability to spectra close to the cluster center, and then drops 
down smoothly toward the  outskirts of the cluster. The scale of this smooth
decrease is provided by the measured distribution of distances in the class.

The algorithm is simple, robust and fast, which makes it ideal to
treat large data sets. It guarantees that similar spectra end up in
the same cluster. Moreover, it is unsupervised since no prior
knowledge of the stellar properties is used,
and the spectra to be classified are the only
information passed on to the algorithm\footnote{
For the sake of comparison, artificial neuronal network 
classifications use a training set that informs the algorithm on the 
existing spectra and spectral types -- see, e.g., 
\citet[][]{2012A&A...538A..76N} and references therein.
}.
These two properties ensure
that the resulting classification is not biased by our (physical)
prejudices, which follows the spirit of a good classification as
advocated by \citet{san05}. Unfortunately, it also has three major
drawbacks. One of them is technical, whereas the other two deal with
the physical interpretation of the classes. The algorithm yields
different classifications depending on the random initialization. This
difficulty is overcome by repeating the classification multiple times,
thus studying the dependence of the final classes on the random seeds.
In addition, our implementation refines the
initialization so that the random seeds are not chosen uniformly but
according to the distribution of points in the classification space
\citep[for details, see ][]{2010ApJ...714..487S}. The second
difficulty  has to do with interpreting the classes as actual clusters
in the classification space, or as parts of larger structures. The
algorithm  does not guarantee that the derived classes correspond to 
actual clusters. However,  one can figure out whether each  particular
class is isolated  or belongs to a larger structure by studying the distances of the 
spectra in the class to the other classes. Well defined classes contain 
stars that are distant from the other classes. The third difficulty refers
to the physical interpretation of the resulting classes, which is
not provided by the algorithm. The physical sense of a particular
class and its cluster center (dubbed in the paper as template spectrum) 
has to be figured out later on. 
Actually, most of this paper is devoted to this task, i.e., to
interpreting in terms  of known stellar physics the  classes resulting from the
k-means classification.

\subsection{Repeatability of the classification}\label{repeatability}
As customary, the dependence of the classification on the random 
initialization was studied by
repeating the classification 100 times, and then comparing the
results. This internal comparison was carried out using three parameters
that we name:
(1) {\em coincidence}, for the percentage of spectra in equivalent
classes,
(2) {\em dispersion}, for the rms fluctuations of the spectra in a
class with respect its cluster center,
and (3) {\em number of classes}, for the number of classes
that contain 99\,\% of the spectra (major classes). 
In order to decide which classes in two different classifications
are {\em equivalent}, we compute the number of stars in common
between each pair of classes formed by one class from one
classification and the second class from the second classification.
The two classes sharing the largest 
number of stars are assumed to be {\em equivalent}.
The same criterion is repeated until all the classes 
of one of the classifications have been paired. 
This criterion maximizes the number of stars sharing
the same class in the two classifications.

\begin{figure}
\includegraphics[width=0.45\textwidth]{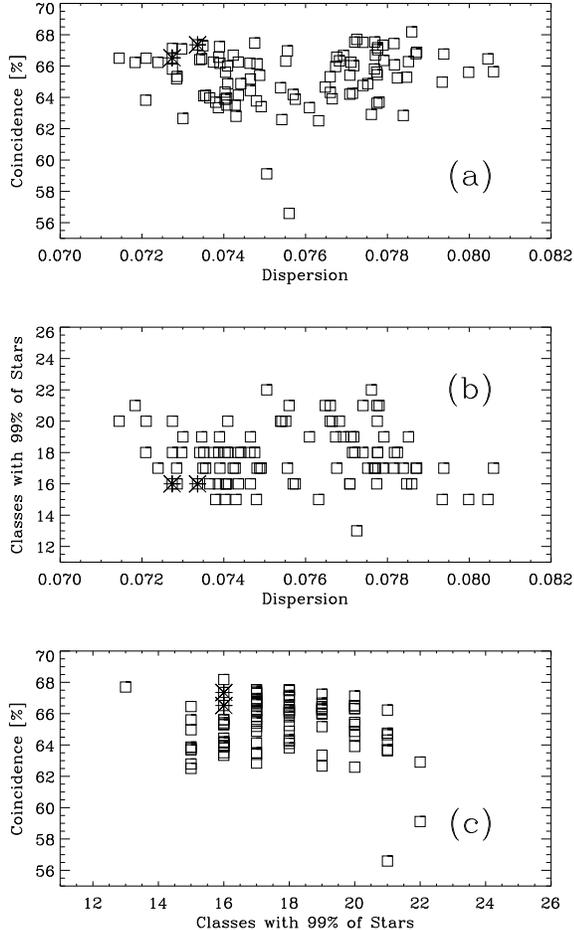}
\caption{
Scatter plots with the three parameters characterizing the 
100 independent classifications of the reference dataset with 
continuum. 
(a) Percentage of stars common to all other 
classifications (coincidence) versus typical scatter 
of the spectra with respect to the class template
(dispersion). 
(b) Number of major classes (i.e., having 99\,\% of the spectra)
versus dispersion. (c) Coincidence versus number of major 
classes.
}
\label{the_class4}
\end{figure}

\begin{figure*}
\includegraphics[width=1.\textwidth,angle=0]{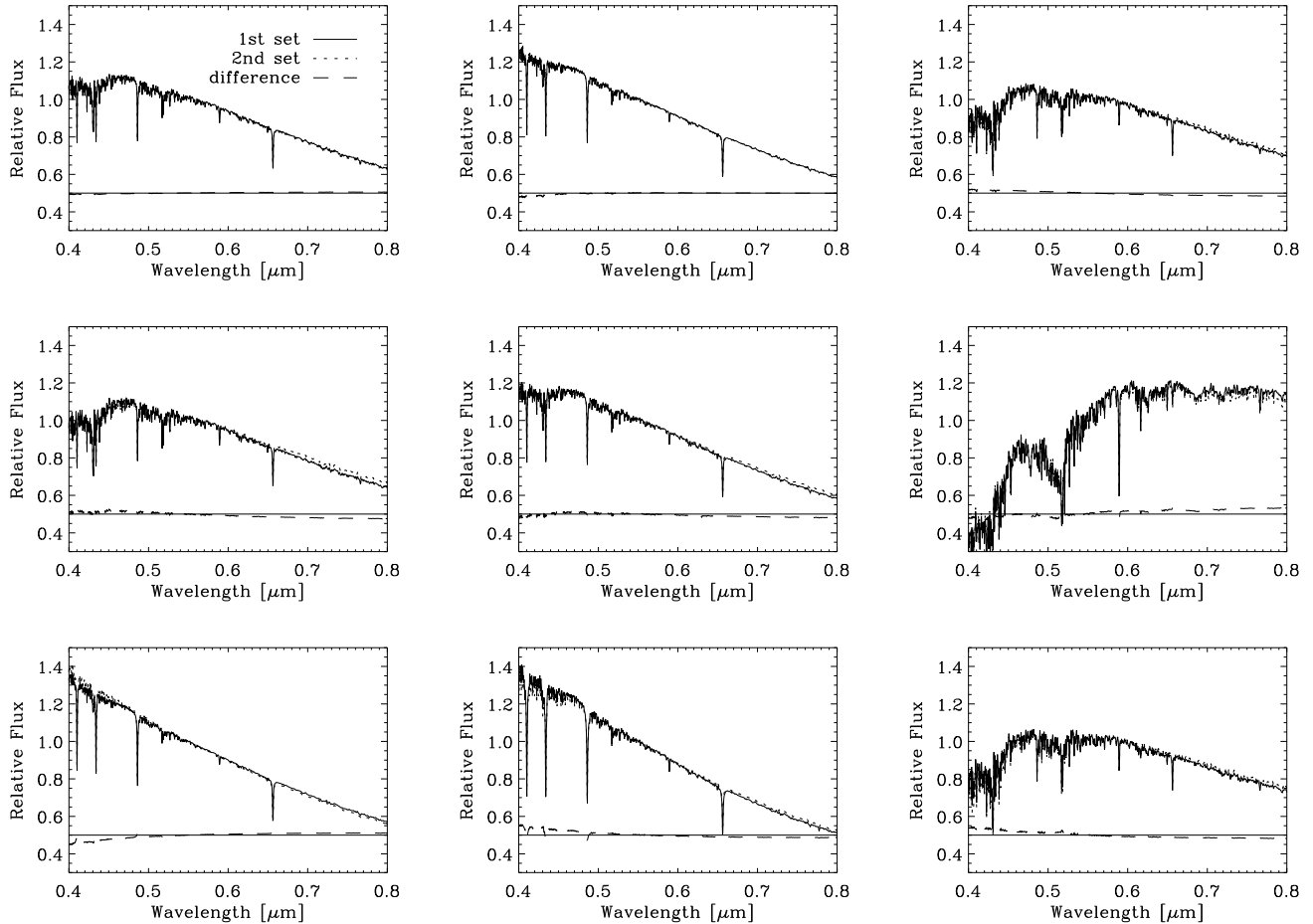}
\caption{
Spectra of several equivalent classes resulting from the 
k-means classification of two disjoint subsets drawn from the 1st auxiliary
set. We show the two template spectra and their difference (uplifted adding 0.5 
so as to fit in the plot); see the inset in the top left panel.
Wavelengths are given in $\mu$m.
}
\label{split1}
\end{figure*}

Figure~\ref{the_class4} shows scatter plots of the three 
diagnostic  parameters corresponding to repeated  classifications of the 
reference dataset (\S~\ref{data}) including continuum. One finds
classifications having between 15 and 20 major classes,
a dispersion in the range 0.07 to 0.08, and a 
mean
coincidence between
62\,\%  and 68\,\%. The ranges in these values are fairly narrow.
The fact that the coincidence is about 65\,\% means that one can
pair the classes of any two classifications, and they will share 
about 65\,\% of the spectra. 
(These apparently low values are discussed below.)
The fact that the dispersion is 
of the order of 0.075 implies that  the differences between 
the class template spectra and the spectra in the class are of the
 order of 7.5\,\% rms. These numbers refer to the reference
dataset including continua (\S~\ref{class_with}) but are
similar to those obtained when spectra without continua 
are used (\S~\ref{class_out}), or when using the auxiliary sets.
In addition to the above tests, we made a numerical experiment 
splitting the 1st auxiliary set  into two randomly chosen
disjoint subsets, which were classified independently.
The differences give an idea on the dependence of the classes
on the particular dataset that is employed. 
The results are summarized in Fig.~\ref{split1}, which shows the 
template spectra of equivalent classes in the two classifications.
(Only nine classes are included, but they are representative of the
general behavior.) The differences between spectra are also plotted, 
and turn out to be of a few percent, i.e., smaller than the 
scatter among the spectra included in each class (the dispersion
of these classifications was of the order of 5\,\%). We have also
computed the colors of the templates and of the individual stars, and
the differences between the colors of the templates 
($\sim$0.025 mag) are smaller than the scatter among
individual stars in a class ($\sim$0.05 mag). 

As mentioned above, repeating the classification 
several times leaves only 60--70\% of the spectra in equivalent 
classes. This issue with k-means is not directly due to the 
random initialization -- the cluster centers in equivalent classes
are very similar, as shown in
Fig.~\ref{split1}  and in \citet[][Sect.~2.1]{2010ApJ...714..487S}. 
It is due to the fact that k-means slices a rather
continuum distribution. Then small changes in the borders  
between clusters produce a significant relocation
of the in-between stars.  
The effect is boosted because the clusters are in a 
high dimensional space  
\citep[see the analytic justification in the appendix of][]{2010ApJ...714..487S}. 
We show in Sect.~\ref{class_with} how some of the resulting 
clusters really have many stars near their borders.

The actual classifications were carried out in parallel using
several workstations.
The  reference data set has a volume of 6.1\,GB,
and the process requires some two CPU hours per single 
classification, or two hundred hours for
studying the effects of the random initialization.
These figures are mentioned to 
stress that the k-mean classification of a sizable data set, 
including  studying initial conditions,  is easily doable  
using standard hardware facilities.

%
\section{The classification: classes and their main properties}\label{classification}

The description given in this section refers to the reference set
defined in \S~\ref{data}, however, the same procedures and analysis
have been repeated with the two auxiliary sets, giving always 
consistent results. 

The classifications use all the 3849 wavelengths equally weighted.
We consider two cases for the analysis.  In the first case,
 the full spectra are used (\S~\ref{class_with}). In the second, 
a running average 193 pixels wide ($\equiv$17,400\,km\,s$^{-1}$) is subtracted from each 
spectrum (\S~\ref{class_out}). This high-pass  filtering removes the 
continuum  but leaves the spectral lines almost untouched.

\subsection{Classes with continuum}\label{class_with}

In order to study the dependence of the classes on the random seeds
inherent to k-means, we carried out 100 independent classifications  
(\S~\ref{repeatability}). 
They are equally valid classifications, but we have to 
choose among them one to be used as  {\em the classification}.  
Taking advantage  of having all these possibilities, we try to selected 
the one that is as representative as possible of all of them,
the spectra in their classes are as similar as possible, and has the smallest 
 number of classes. In the parlance used in \S~\ref{repeatability}, we 
try to select a classification having large coincidence,  
small dispersion and few classes. 
The scatter plots for these three parameters among the 100 independent 
classifications are shown in Fig.~\ref{the_class4}. Asking the coincidence 
to be larger than 66\,\%, the dispersion to be smaller than 0.074,
and the major classes to be fewer than 17, one finds only two 
classifications. Those are represented as asterisks in Fig.~\ref{the_class4}.
For lacking  a better criterion, we choose one of them at random.
Its coincidence is 67\,\%, its  dispersion 0.073, and  it has
16 major classes (26 in total, but some seem to correspond 
to failures of the SDSS pipeline, as we explain later on).

\begin{figure*}
\includegraphics[width=0.89\textwidth,angle=0]{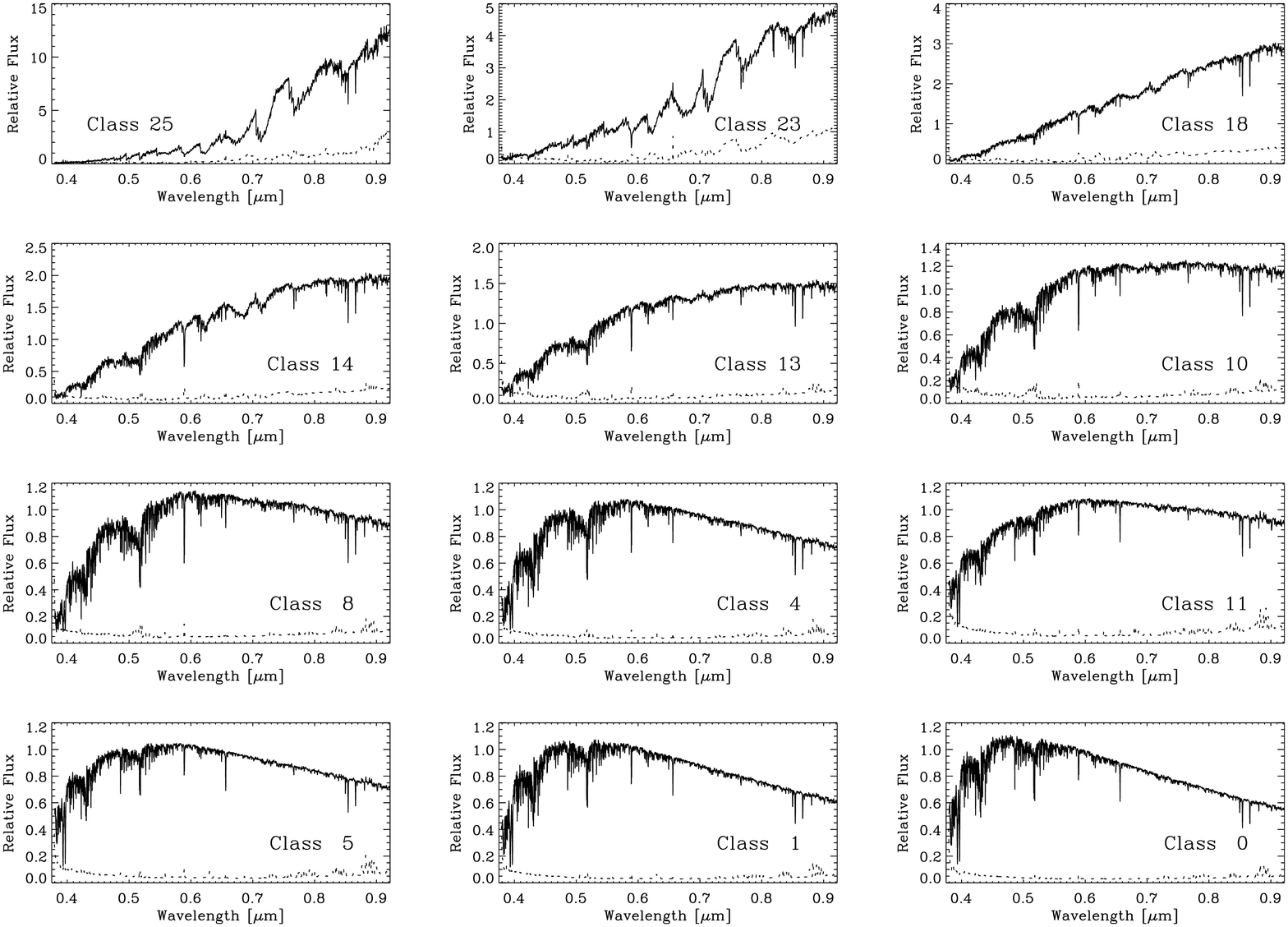}
\includegraphics[width=0.90\textwidth,angle=0]{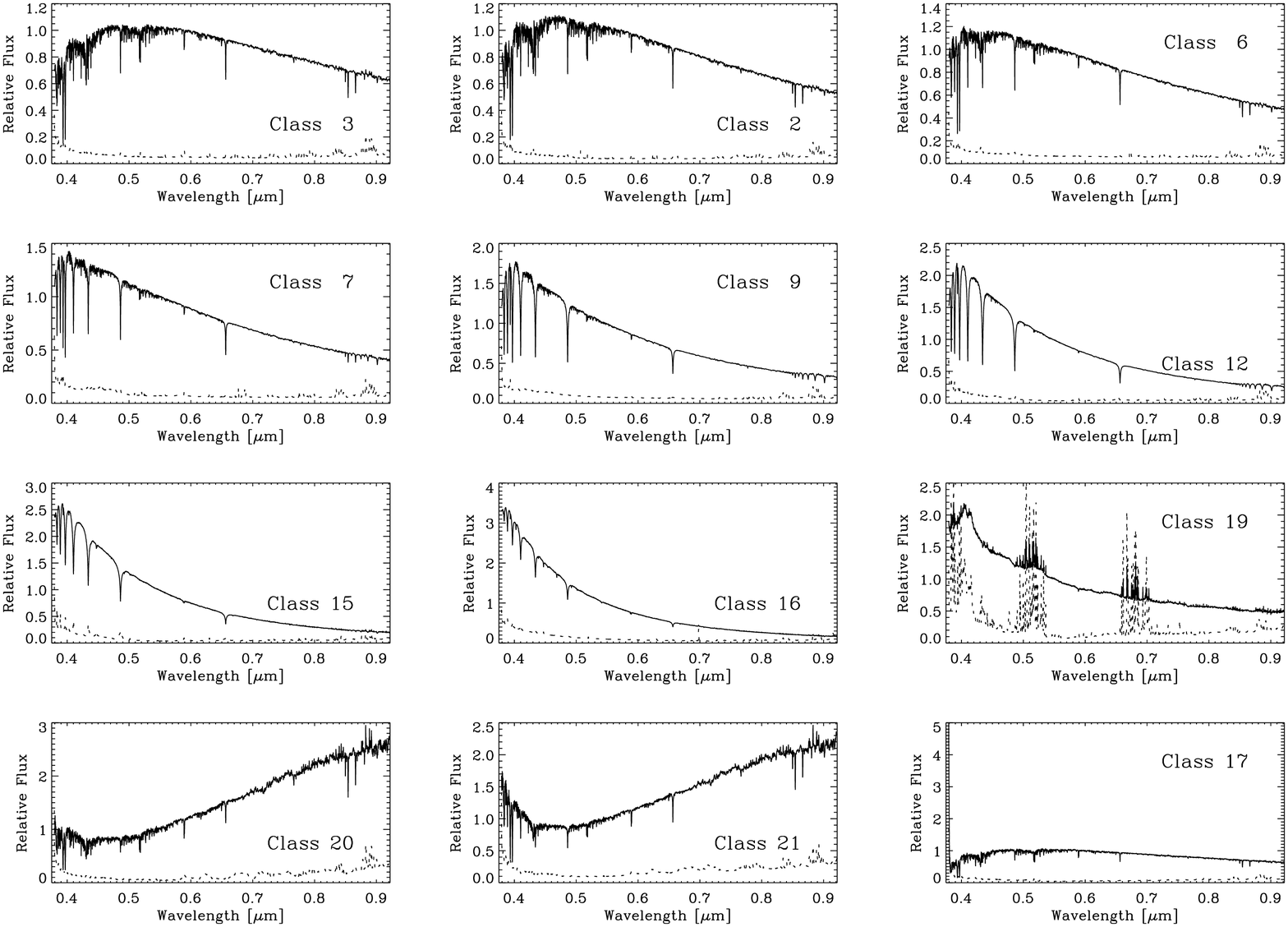}
\caption{
Template spectra of all classes in the classification that includes 
continuum  (the solid lines).
The spectra are normalized to the intensity at some 5500\,\AA,
and the individual plots are scaled from minimum to maximum.
The classes are identified in the insets, and they have been ordered
from red to blue (from left to right and from  top to bottom)
following Fig.~\ref{the_class_temp4otro}.
This order breaks down with the abnormal classes 19, 20, 21 and 17, 
that are shown at the end of the sequence.
Classes 22 and 24 are not included,
since they correspond to failures in the reduction pipeline and 
are similar to class~17. 
The panels also include the intra-class 
standard deviation, that quantifies the dispersion among the 
spectra included in the class (the dotted lines). 
Wavelengths are given in $\mu$m. 
}
\label{the_class_temp4}
\end{figure*}

\begin{figure}
\includegraphics[width=0.5\textwidth,angle=0]{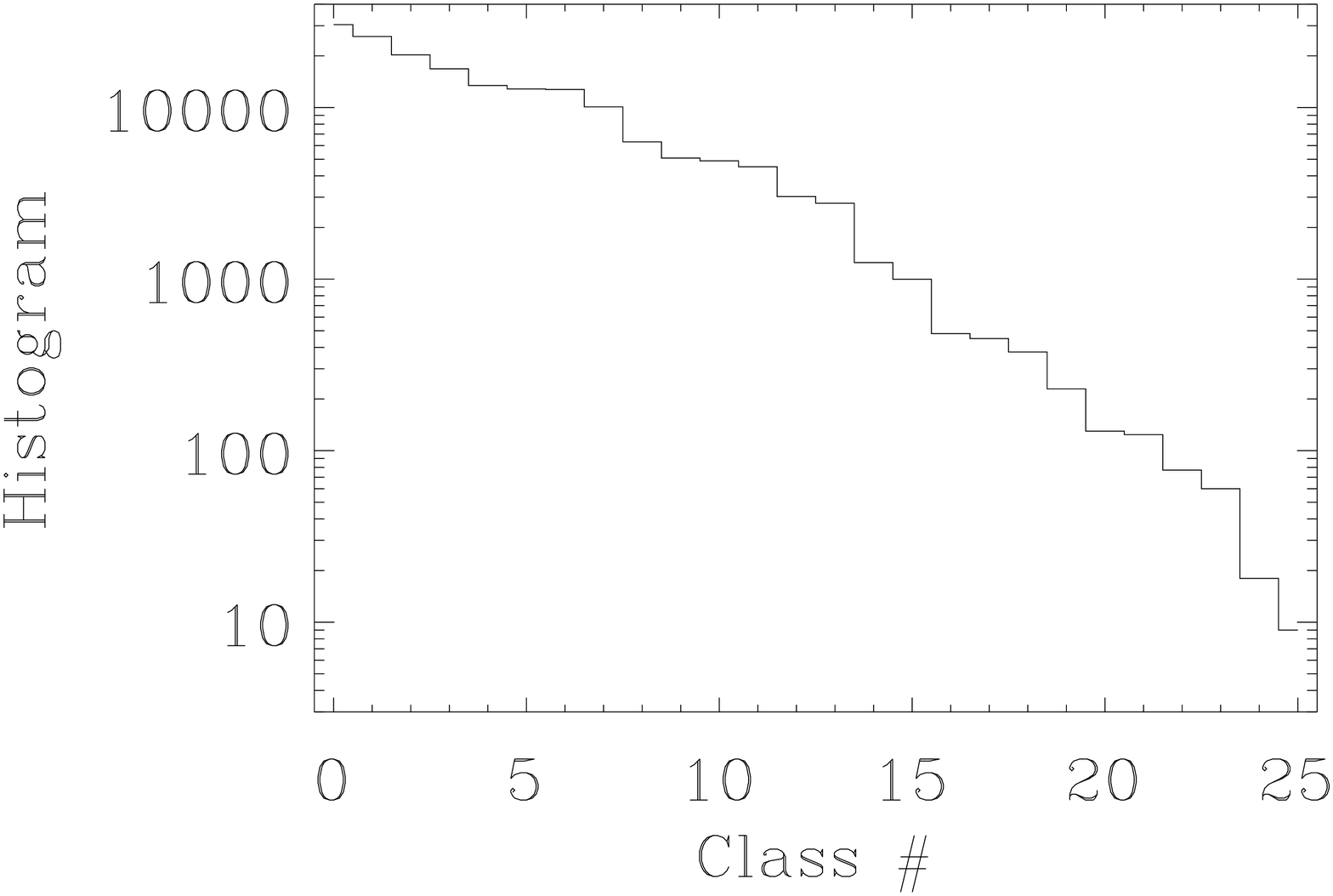}
\caption{
Histogram with the number of spectra in each class as derived
from the k-means classification of stellar spectra that includes 
continuum.
The class number has been assigned
according to the number of stars in the 
class, being class~0 the class with the largest number of
elements.
}
\label{the_class4_his}
\end{figure}

\begin{figure}
\includegraphics[angle=0,width=0.5\textwidth]{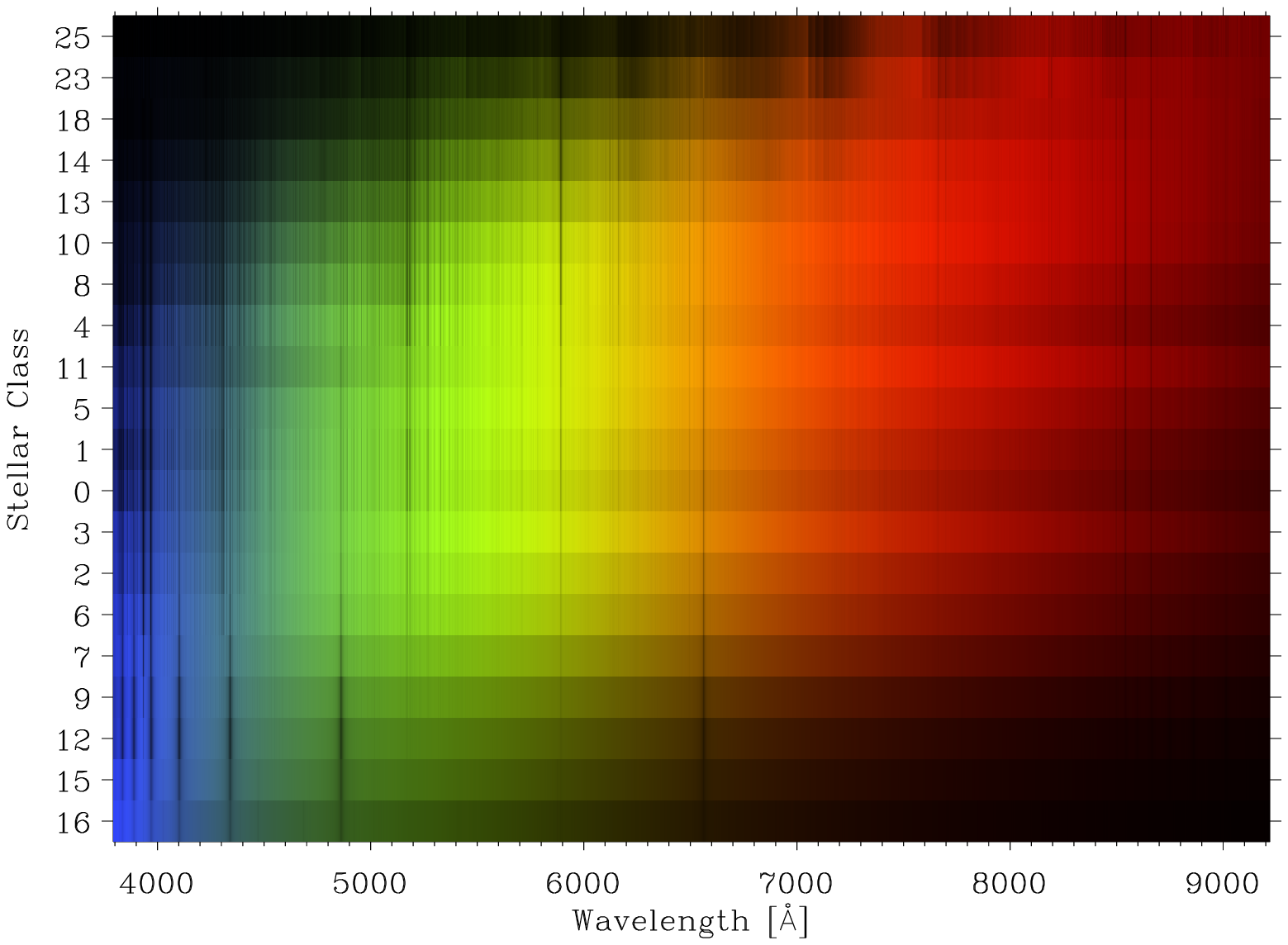}
\caption{
Composite image with the templates spectra of all the classes in the k-means 
classification of spectra with continuum. They have been ordered so that the
image looks  smooth. All spectra are normalized to their maximum intensities. 
The color palette tries to mimic the human eye sensitivity. The classes that correspond to 
failures of the pipeline (classes 17, 22 and  24) and  binary systems (classes 20 and 21)
are excluded.
}
\label{the_class_temp4otro}
\end{figure}

The average of all the spectra in the classes (i.e., the cluster centers
or the cluster templates) are shown 
as solid lines
in Fig.~\ref{the_class_temp4}.
The figure also includes the standard deviation among all the spectra 
in each class (the dotted line), which quantifies  
the intra-class dispersion.  
The number of stars in each class is represented as  an histogram 
in Fig.~\ref{the_class4_his}. It shows that the classes have 
been numbered according to the  stars than contain, being 
class~0 the most numerous, class~1 the second most numerous,
and so on and so forth. 
Since the template spectra come from averaging 
thousands of individual spectra, they have extremely high
signal-to-noise ratios -- from 200 to 2000 depending on the
number of spectra in the class.
The spectra of classes~22 and 24 are not included in 
Fig.~\ref{the_class_temp4}. 
They collect faulty spectra that  are  similar to class~17  
(see the template spectrum in Fig.~\ref{the_class_temp4}, 
that has a large unphysical spike at the bluest wavelength).
%
The templates are also represented in Fig.~\ref{the_class_temp4otro} as a stack-plot
ordered so that the image looks as smooth as possible. This image excludes 
those classes that are failures of the pipeline (classes 17, 22 and  24) and  
binary systems (classes 20 and 21; see below).

\begin{figure}
\includegraphics[angle=0,width=0.5\textwidth]{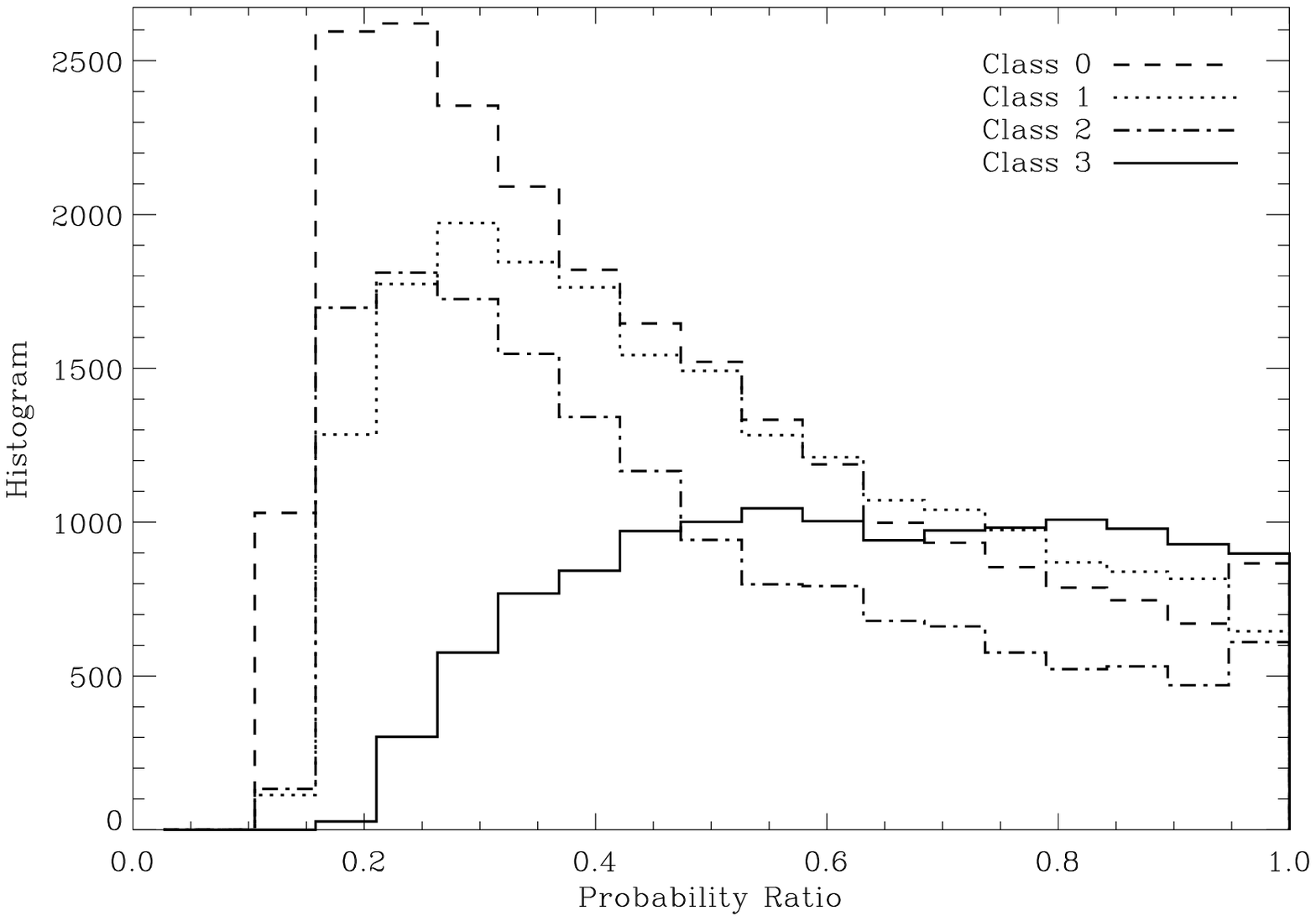}
\caption{
Histograms of the ratio between probabilities of belonging to the
2nd nearest cluster and to the assigned cluster for 
 stars included in classes 0 to 3 (as  the inset indicates).
The abscissae are by definition bound between 0 and 1. Histograms
peaking at low ratio characterize well defined classes (e.g., class~0)
whereas histograms with large counts at large ratios indicate
a fuzzy class (e.g., class~3).  
}
\label{groups_stars_pub}
\end{figure}

As we discuss in Sect.~\ref{algorithm},  k-means does not guarantee 
the inferred classes to be  real clusters in the classification space. 
They may be parts of larger structures that have been sliced by the algorithm. 
A way to study whether the classes are isolated was explored in 
\citet{2010ApJ...714..487S}, and it is used here too. 
One can estimate the probability that each particular star belongs to the 
class it was assigned to.  It depends on how far from the 
cluster center the star is  as compared to the other members in its class. 
Similarly, one can 
estimate the probability of the star belonging to any other class.  
Well defined clusters will have most of
their elements with a probability of belonging to any other 
cluster significantly smaller than the probability of belonging to the cluster.
Figure~\ref{groups_stars_pub} shows histograms of the ratios
between probabilities of belonging to the 2nd nearest cluster and to the 
assigned cluster for  a few representative classes.
There are classes where the histogram peaks at low ratios, thus
indicating a well defined structure in the classification space
(e.g., class~0 in Fig.~\ref{groups_stars_pub}).
Conversely, other classes present a rather flat histogram indicating
a dispersed structure (e.g., class~3 in Fig.~\ref{groups_stars_pub}).
Classes 3, 5, 11 and 19 represent spread-out classes, whereas 
the rest are clustered classes. 
Note, however, that even the histograms of 
well defined clusters have a significant tail towards large ratios, indicating the 
presence of many stars in the boundaries between clusters. Those stars are partly 
responsible for the non-uniqueness of the classification studied in Sect.~\ref{repeatability}.

\begin{figure}
\includegraphics[width=0.5\textwidth,angle=0]{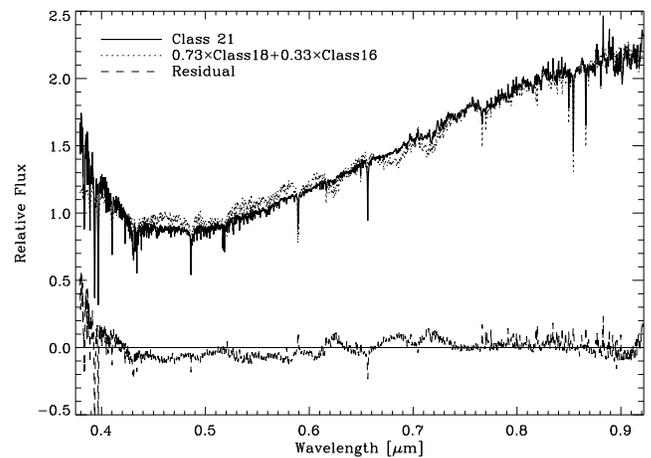}
\caption{
The class 21 template spectrum (the solid line), with upturns in the blue and the red, 
seems to be the composite spectrum of a stellar system. 
Assuming it to be binary, the
best fit (the dotted line) is obtained as a linear superposition of a 
WD (class~16) and a late type giant (class~18). The fit is not perfect;
the difference between class~21 and the best fit is shown as the dashed line.
}
\label{composite1}
\end{figure}

The physical interpretation of some of the classes is relatively 
straightforward. 
Classes 20 and 21, with upturns in both the blue and the red, are 
most likely composite 
spectra of  systems with two (or more) stars with very different effective 
temperatures ($T_{\rm eff}$). 
They can be gravitationally bounded stellar systems, or stars that 
happen to be along the line of sight.
The luminosity of the stars that contribute 
to the combined spectrum has to be similar, therefore,
in case of binary systems both stars cannot be in the main sequence
because the hot star would outshine any cold companion. 
One common possibility is a hot  white dwarf (WD) and 
a cold dwarf or giant,
and this is indeed the conclusion reached when trying to reproduce the 
templates of classes~20 and 21 as a linear superposition of templates of
two other classes. The best fit is obtained combining classes 16 and 18, as 
shown in Fig.~\ref{composite1}. 
As we discuss later on, class~16 contains WDs, and class~18 corresponds 
to K-type giants.
We note that the templates of classes~20 and 21 resemble spectra
of post-common envelope binaries, as identified and studied
using SDSS data
\citep[e.g.,][]{2007MNRAS.382.1377R,2008A&A...484..441S,2008MNRAS.390.1635R}.
Some other classes are really awkward, and so difficult to interpret unless they
are associated with failures in the reduction pipeline (e.g., classes~17 and 19).

\begin{figure}
\includegraphics[width=0.5\textwidth,angle=0]{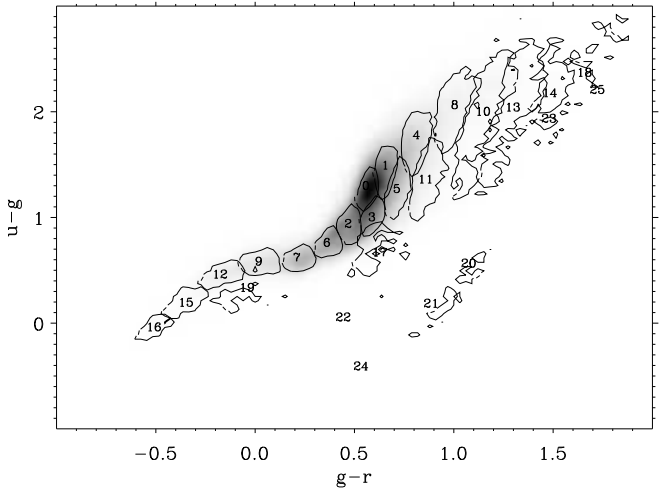}
\caption{
$u-g$ vs $g-r$ plot for the full set (the image in the background)
and for the different classes separately. The contours show the region
with 68\% of the spectra in the class, and have been
labeled with the corresponding class number.
This plot corresponds to the full spectrum classification.
Classes 17, 19,  22, and 24 seem to be failures of the 
reduction pipeline. Classes~20 and 21 are binary or multiple systems with 
spectra blue in the blue filters (i.e., $u-g$) and red in the red 
filters (i.e., $g-r$).
The rest of classes form a 1D sequence that bifurcates at $g-r\simeq 0.5$
(at classes 0 and 3).    
}
\label{the_class_col4}
\end{figure}

Figure~\ref{the_class_col4} shows an image with the distribution
of $u-g$ vs $g-r$ colors of the full set,  with the individual classes
overlaid as contours containing 68\% of the stars.  
The colors have been derived
from the spectra using the transmission bandpasses of the 
broad-band SDSS filters. Note that the classification is basically a color classification.
Disregarding classes gathering  failures of the pipeline
(classes 17, 19,  22, and 24) and multiple systems (classes~20 and 21),
the k-means classification of stellar spectra with continuum seems to
separate stars according to
their position on the color-color plot. The classes  form a one-dimensional
set in the diagram, with a bifurcation at $g-r\simeq 0.5$.
The bifurcation separates dwarf stars (on top)
from giant stars, a result further discussed below. 
We note, in passing, that multiple systems are
well separated  in this color-color plot and, therefore, it
can be used to select them.

\begin{figure}
\includegraphics[width=0.5\textwidth]{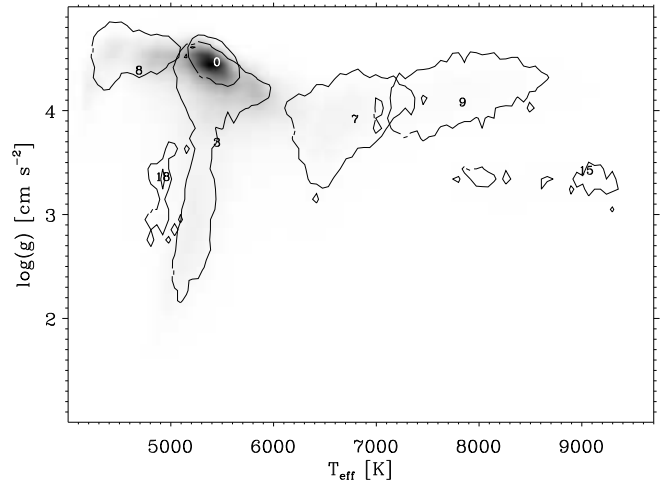}
\caption{$\log({\rm g})$~vs~$T_{\rm eff}$ for 
different classes resulting from the classification
of stellar spectra with continuum. The 
contours containing 68\,\%\ of the stars in the class.
Only classes~0, 3, 7, 8, 9, 15 and 18 are included
to avoid cluttering the figure.  The class numbers 
have been placed at the location of the mean of the
corresponding distribution.
}
\label{density4}
\end{figure}

Main sequence stars
have $\log({\rm g})$ larger than 3.8 (with the 
surface gravity g in cm\,s$^{-2}$), which is even larger ($\log({\rm g})> 4$) 
for stars with $T_{\rm eff} < 9000\,$K \citep[e.g.][]{2000asqu.book..381D}.
Figure~\ref{density4} shows the two-dimensional  distribution
of $\log({\rm g})$~vs~$T_{\rm eff}$, together with contours
with the region containing 68\,\% of the stars in
the class.  The gravities and effective temperatures of individual
stars have been taken from the SSPP (parameters labeled ADOP), as we
explain in Sect.~\ref{data}. The plot does not include all the classes
since many of them overlap and would clutter the figure. Only those classes
relevant for our argumentation are included, in particular
classes~0 and 3 have similar $T_{\rm eff}$ but are parts of the two
different branches of the color sequence (see Fig.~\ref{the_class_col4}). 
Note that class~0 gathers only main sequence stars ($\log({\rm g}) > 4$)
whereas most class~3 targets are giants.
Something similar occurs with the pairs of classes 1 and 5, and  
8 and 11. Classes 0, 1, 4 and 8  contain only main sequence stars 
(see classes~0 and 8 in Fig.~\ref{density4}).
Classes 14 and 18 contain only giant
stars  (see class~18 in Fig.~\ref{density4}). Several classes do not
 have enough valid $T_{\rm eff}$ and 
$\log({\rm g})$ to know their location in the 
 $\log({\rm g})$~vs~$T_{\rm eff}$ plot, including the classes
with faults plus classes 16 and 25. Class 25 is a minor class
with few elements, but class~16 is not. The lack of
 effective temperatures and gravities for class~16 seems
to be associated with the fact that it collects WDs, for 
which no proper physical data are provided by 
the SSPP 
\citep[but see, e.g.,][]{2006ApJS..167...40E,2010AIPC.1273..156K,2011ApJ...730..128T}.
We note that classes 20 and 21, multiple systems whose spectra combine
hot and cold components (see Figs.~\ref{the_class_temp4} and \ref{composite1}),
appear in Fig.~\ref{density4} as main sequence stars  with  
$T_{\rm eff}$ between 5000\,K and 6000\,K. They are also metal rich  systems according to the plot 
discussed in the next paragraph.

\begin{figure}
\includegraphics[width=0.5\textwidth]{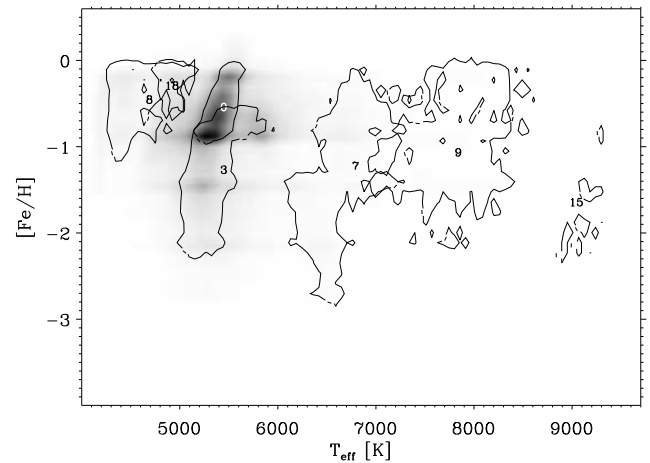}
\caption{[Fe/H]~vs~$T_{\rm eff}$ for some representative
classes resulting from the classification of spectra including continuum.
The represented classes are the same as those in Fig.~\ref{density4}. 
The contours embrace 68\,\%\ of the stars in the class, and the class 
numbers have been placed at the mean point of the corresponding
distributions.   
}
\label{fe_vs_teff4}
\end{figure}

Figure~\ref{fe_vs_teff4} shows the two-dimensional distribution of [Fe/H]~vs~$T_{\rm eff}$, 
together with contours with the region containing 68\,\% of the stars in
the class. As the rest of physical parameters of stars,
the metallicity [Fe/H] comes from the SSPP. 
The figure shows how the classes often contain both high and low metallicity 
stars. If the threshold between low and high metallicities is set at one tenth of
the solar value (i.e.,  [Fe/H]$= -1$), the classes that contain
only high metallicity stars are 0, 1, 4, 8, 14, and 18. Similarly, 
classes 12  and 15 include only low metallicity stars. Some of these
classes are included in  Fig.~\ref{fe_vs_teff4}.
Class 18 contains low gravity, low temperature high metallicity stars -- probably
K giants. 
Class 15 contains low gravity high temperature low metallicity stars.

\begin{figure*}
\includegraphics[width=0.8\textwidth]{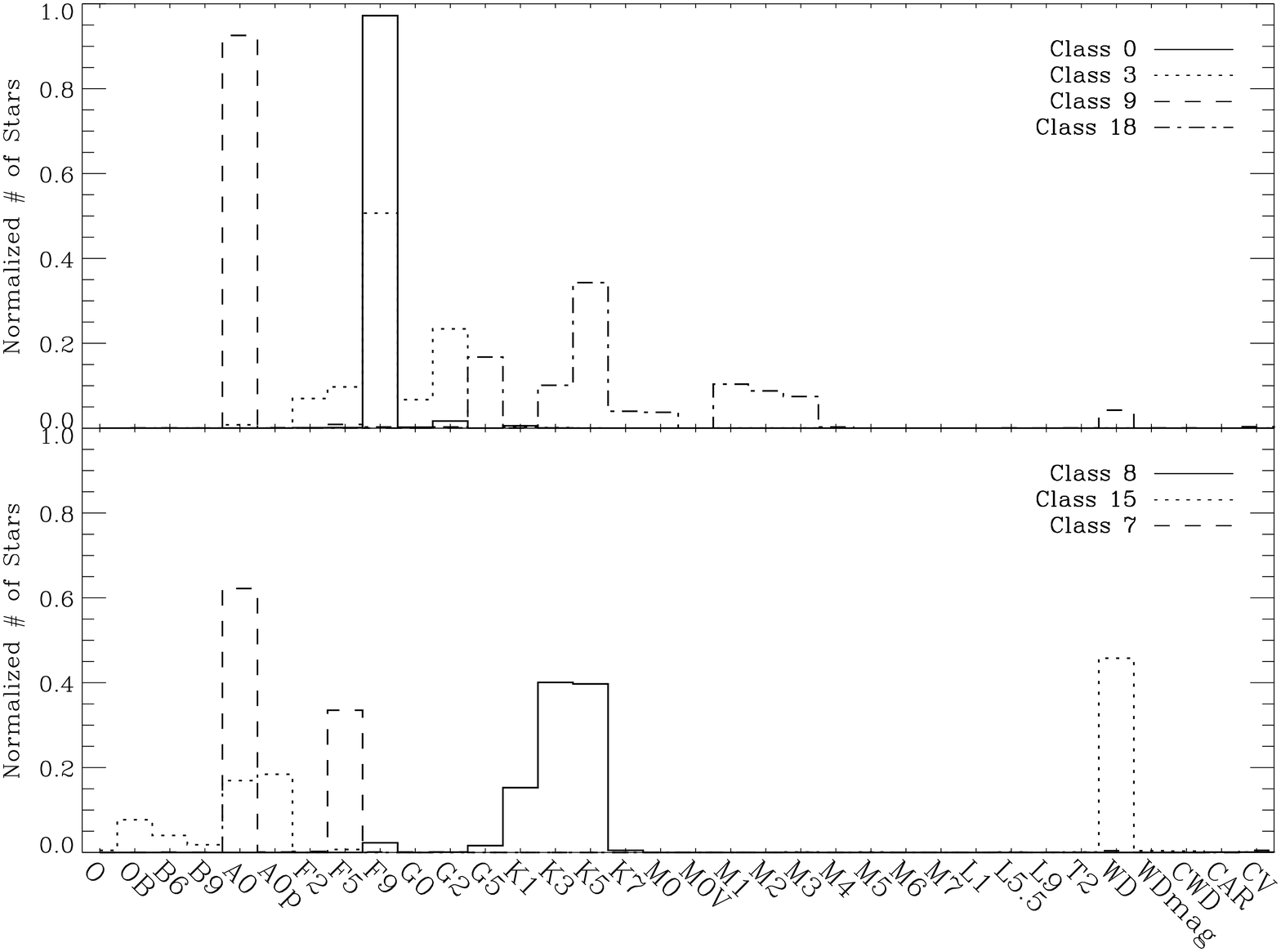}
\caption{
Histograms of the distribution of MK types of selected k-means classes
corresponding to spectra with continuum. (The classes 
are those chosen for Figs.~\ref{density4}
and  \ref{fe_vs_teff4}.) Splitting the histograms
into two panels avoids overcrowding the figure. 
The MK types shown here and in Fig.~\ref{original_classes2}
are directly comparable. The histograms have been 
normalized to one, including objects for which the 
MK type is not available. 
}
\label{mkclasses4}
\end{figure*}

The k-means classes do not exactly coincide with 
the classical MK types assigned to the  stars by 
\citet[][]{2008AJ....136.2050L}
(see Sect.~\ref{data}).  
Figure~\ref{mkclasses4} presents the histogram of
MK types corresponding to the k-means classes in 
Figs.~\ref{density4} and \ref{fe_vs_teff4}. As the histograms show,
most classes can be ascribed to a single MK type or to a narrow range
of them (e.g., class~0, 8 and 9). However, the spread in MK
types is sometimes large (e.g., class~18), becoming  
extreme in the bluest classes (e.g., class~15), which often group 
A-type stars (mainly on the horizontal branch) 
with WD.   

To recapitulate, the k-means classification of spectra with continuum seems to be
basically 
a color classification. The stellar colors are driven mostly by the overall continuum
shape, therefore, the k-mean classes separate stars according to 
their continua.

\subsection{Classes with continuum removed}\label{class_out}

As explained in the previous section, the k-means classification of  stellar
spectra with continuum is essentially  a color classification. Since the 
colors are dictated mostly by the continuum, the classification is
driven by the shape of the continuum. 
Dust extinction and errors in the spectro-photometric calibrations
corrupt continua but not so much spectral lines, which retain the 
information on the stellar properties.  
In order to study the potential of k-means to identify
and separate spectra according to  subtle spectral-line differences, we repeated
the classification using spectra without continuum. Explicitly,
the spectra to be classified are the original spectra after removal
of a running mean filter 193 pixels wide, which corresponds to
170\,\AA\ in the blue and 400\,\AA\ in the red.
The width of the numerical filter was determined as a trade off to be
broader than most spectral features, yet narrow enough to be representative 
of the local continua.  Because SDSS spectra are sampled in log wavelength, our
constant width in pixels represents a single Doppler broadening 
of the order of 17,400 km\,s$^{-1}$. 

\begin{figure}
\includegraphics[width=0.5\textwidth,angle=0]{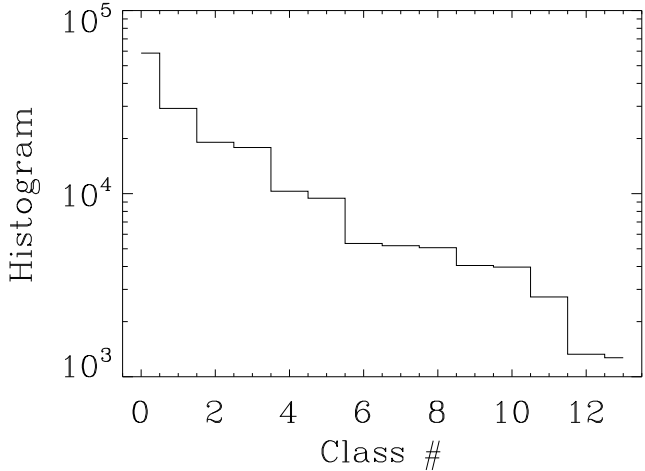}
\caption{
Histogram with the number of spectra in each class for the
classification of spectra without continuum.
The class number has been assigned
according to the number of stars in the 
class, being 0 the class with the largest number of members.
}
\label{the_class6_his}
\end{figure}

\begin{figure*}
\includegraphics[width=1\textwidth,angle=0]{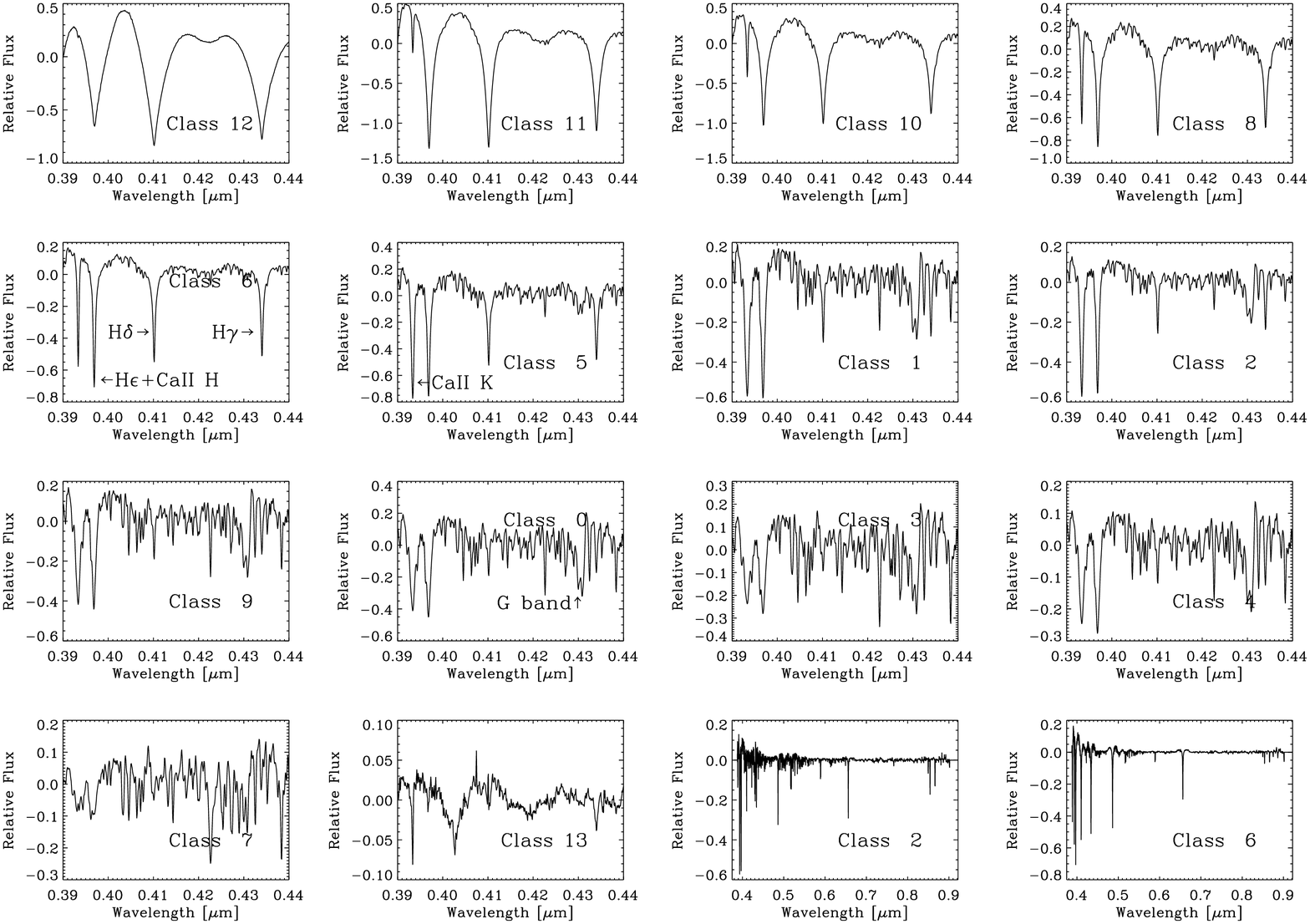}
\caption{
Template spectra of all the classes in the classification
with the continuum removed. We just show a small portion in the 
blue range, otherwise it is impossible to appreciate individual 
spectral lines. The exception is given by the two panels
at the lower right corner, where classes 2 and 6 are repeated
in their full spectral range.   The class numbers are given as
insets, an some of the main spectral features in the region are
also labelled. 
The classes have been ordered following 
Fig.~\ref{the_class_temp6otro}
(from left to right and from top to bottom).
}
\label{the_class_temp6}
\end{figure*}

\begin{figure}
\includegraphics[angle=0,width=0.5\textwidth]{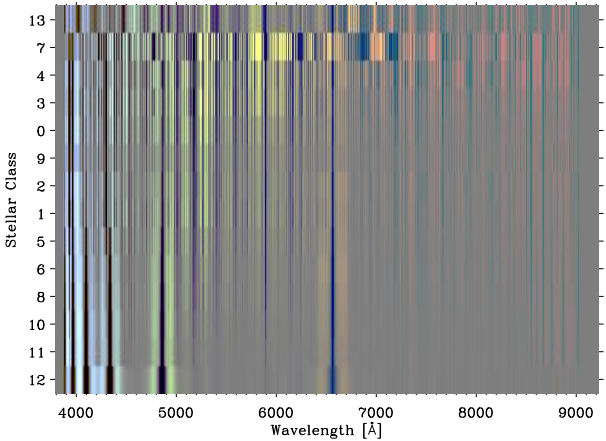}
\caption{
Template spectra of all the classes corresponding to the 
classification without continuum. They have been ordered
so that the image looks smooth.
Class~13 on top seems to collect spectra with instrumental problems.
}
\label{the_class_temp6otro}
\end{figure}

\begin{figure*}
\includegraphics[angle=0,width=1\textwidth]{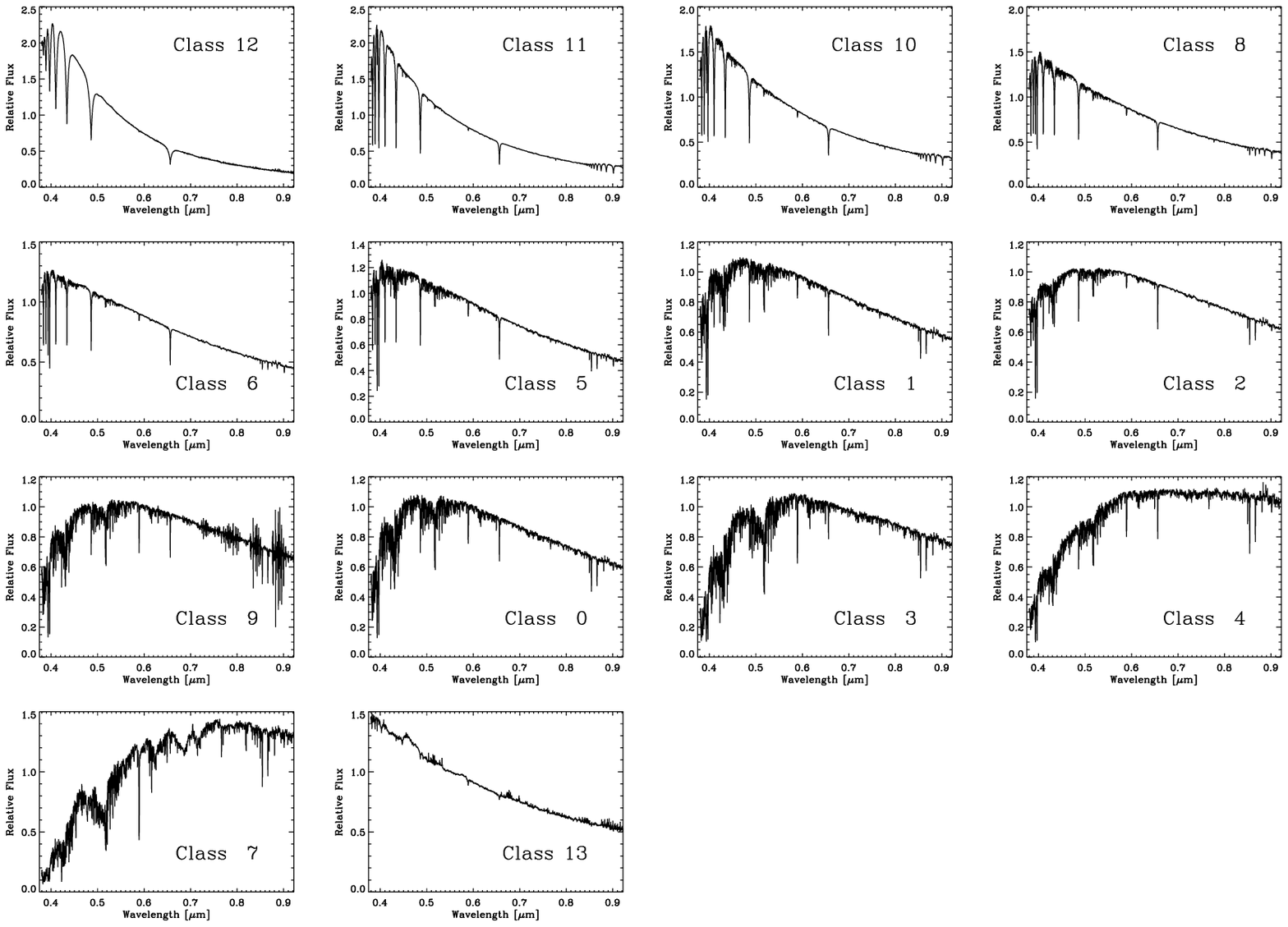}
\caption{
Average spectra of the classes resulting from classifying 
the spectra without their continua. Even though the continuum was removed
for classification, it has  been included in this average. 
Note the suspicious shape of class~13,
and the emission line features of class~9 at some 0.9~$\mu$m.   
The classes have been ordered following 
Fig.~\ref{the_class_temp6otro}, as in Fig.~\ref{the_class_temp6}.
}
\label{the_class_temp6mas}
\end{figure*}

The procedure leading to the classification is similar to that used 
for the full spectra described in the previous Sect.~\ref{class_with}. 
The {\em dispersion}, {\em coincidence}, and 
{\em number of major classes} were used to select one among 100 
independent initializations. The selection criteria try to make this classification  
as representative of the rest as possible. The selected class has 
${\rm coincidence}=75.7\,$\%, ${\rm dispersion}=0.051$,
13~mayor classes, and 1~minor class. Note that the coincidence is 
larger than that for the classification with continuum, and the dispersion 
and number of classes smaller.
The number of stars in each class is shown in Fig.~\ref{the_class6_his}.
As we did for the classes resulting from classifying the spectra
with continuum (Sect.~\ref{class_with}), the new classes are also named 
class~0, class~1, and so on, with the number increasing as the 
elements in the class decrease.
The average spectra of all the stars in each class are shown in 
Fig.~\ref{the_class_temp6}. We just show a small portion of the blue
spectrum where individual spectral lines can be appreciated.
(Two spectra spanning the full spectral range are shown for illustration,
but then it is impossible to appreciate details of the lines.)
Note that each template spectrum is
the average of thousands of individual spectra, so all the 
small noise-looking wiggles  are real spectral features. 
The full templates are shown as a stack-plot image in Fig.~\ref{the_class_temp6otro}
 (c.f. Fig.~\ref{the_class_temp4otro}).
They have been ordered so that the sequence looks as smooth as possible.
The Balmer lines,  which are the only ones present in class~12,
decrease in strength as one moves up in the image.  The 
conspicuous molecular bands of TiO  are present only 
in class~7.  Figure~\ref{the_class_temp6mas} also shows the 
average spectra of the classes, but this  average was computed using 
the original spectra with their continua intact.  The comparison of these spectra
with those corresponding to the classification including 
continuum  (Fig.~\ref{the_class_temp4}) renders a few
differences. First, the faulty classes with a fake emission peak 
in the blue  (e.g., class~17  in Fig.~\ref{the_class_temp4}) 
have disappeared.  This is a side-effect of removing the continuum  from 
the spectra, which in our implementation blacks out the 193 pixels
in the extreme wavelengths, thus removing the problem.
Second, there is a new class that shows spectra with 
emission lines (class~9), which most probably are not 
real but poorly corrected sky lines at the wavelengths of the
Ca II IR triplet \citep[see., e.g.,][]{2008AJ....136.2050L}.
Finally, the spectra corresponding to binary systems do not form
a separate class, so that they have to show up as outliers of the
classification (Sect.~\ref{outliers}).    

\begin{figure}
\includegraphics[width=0.5\textwidth,angle=0]{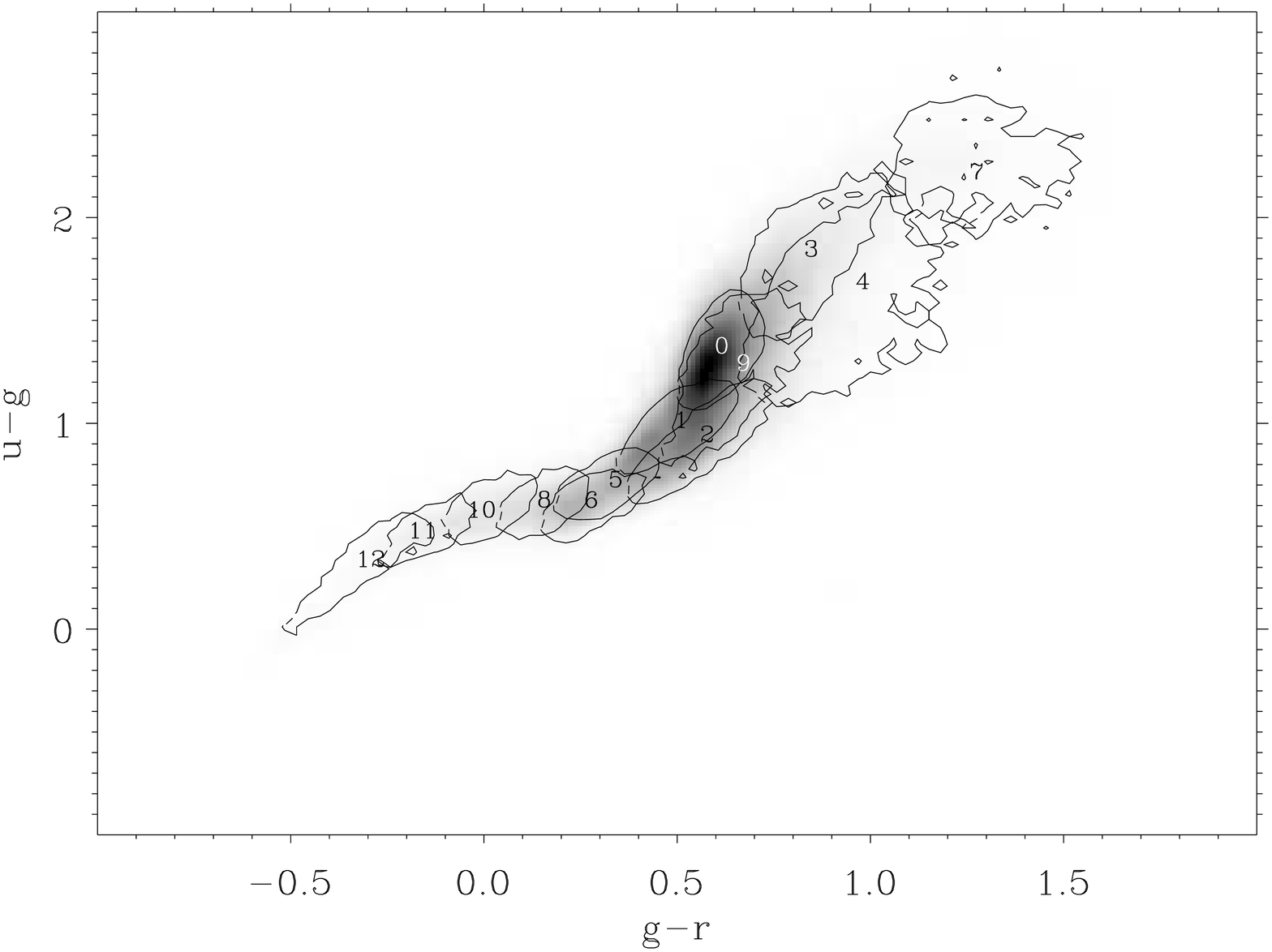}
\caption{
$u-g$ vs $g-r$ plot for the reference set of stellar spectra
(background image) and for the different classes separately. The contours 
show the region with 68\% of the spectra in the class, and  the
centroid of each distribution has been labeled with the class number 
it belongs to. This plot corresponds to the classification where
the continuum was removed. 
}
\label{the_class_col6}
\end{figure}

\begin{figure}
\includegraphics[angle=0,width=0.5\textwidth,origin=c]{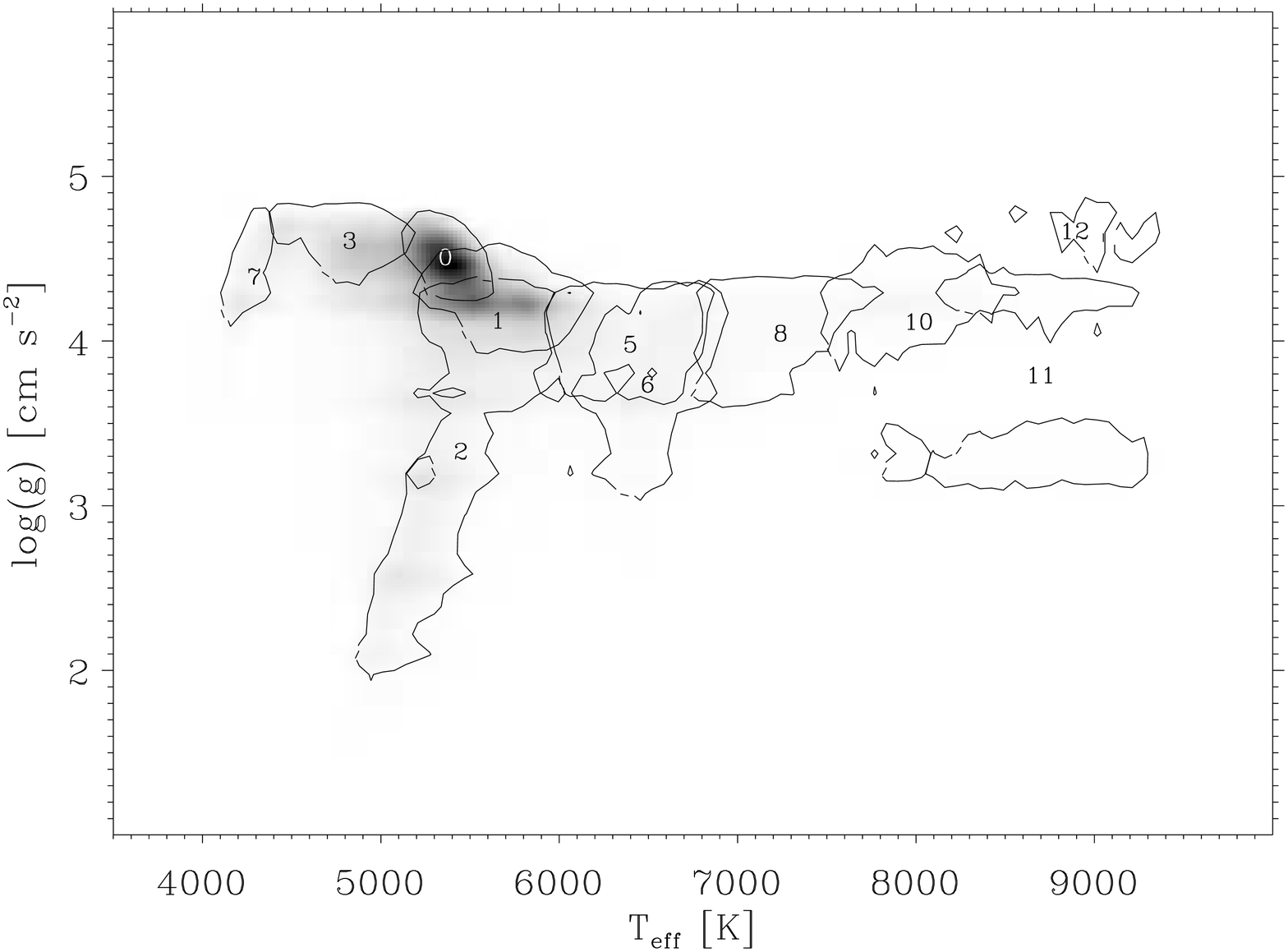}
\caption{$\log({\rm g})$~vs~$T_{\rm eff}$ for a number of selected classes resulting 
from the classification of the stellar spectra with the continuum removed. 
The image shows the full set, and the contours indicate the regions
 containing 68\,\%\ 
of the stars in the class. The class numbers appear at the centroid of the
distributions.   
The distribution of class~11 has two separated peaks
at $\log({\rm g})\sim3.2$ and 4.2, so that the corresponding label appears
between them.
}
\label{density6}
\end{figure}

The classes are shown in the $u-g$ vs $g-r$ color plot in Fig.~\ref{the_class_col6}.
They overlap more than the classes inferred when the continuum is included;  see
Fig.~\ref{the_class_col4}. 
There are several conclusions to be drawn from the comparison of these two figures.
The continua influence the classification or, in other words, the classifications with and 
without continua do not fully agree. However, most classes can be 
viewed as mergers of classes with continuum. Even if  the continuum is not 
included for classifying, the different classes have different colors --  
classes 0 and 9  represent an exception since they overlap in the
color-color plot (see Fig.~\ref{the_class_col6}).
The color-color plot also shows two parallel sequences that
split at $g-r\sim 0.2$ (or at  class~6). As it happens with the
classification including continuum, the upper branch corresponds
to main sequence stars whereas the lower branch includes giants.
This separation by stellar size is more clear in  Fig.~\ref{density6},
which shows the 2D distribution
of $\log({\rm g})$~vs~$T_{\rm eff}$ for
the stars in a number 
of selected classes -- the classes in the 
lower branch of Fig.~\ref{the_class_col6}  include low gravity 
stars in Fig.~\ref{density6} (see classes~6 and 2).  
\begin{figure}
\includegraphics[angle=0,width=0.5\textwidth]{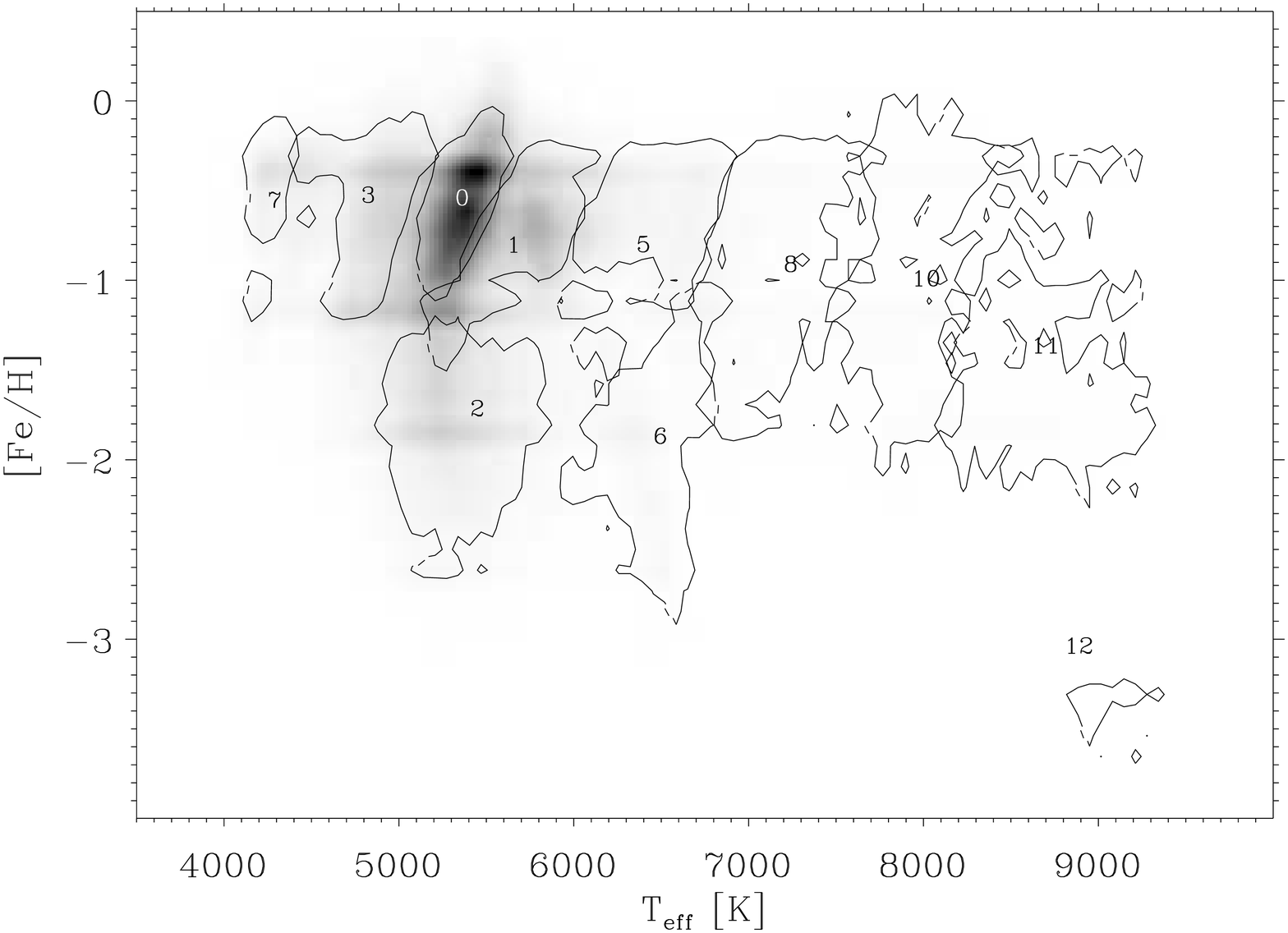}
\caption{[Fe/H]~vs~$T_{\rm eff}$ for the 
different classes resulting from the classification
of the SEGUE stars
with their continua removed. The image displays the histogram of the full
set, whereas the contours mark the regions containing 68\,\%\ of the stars in the 
class.
Class numbers are located at the center of the corresponding distribution.
}
\label{fe_vs_teff6}
\end{figure}
Figure~\ref{fe_vs_teff6} shows the 2D distribution
of [Fe/H]~vs~$T_{\rm eff}$ for the full set of stars, together with 
contours with the region containing 68\,\% of the stars in
selected classes. Note how well separated are the classes in this
plot, in contrast with the overlap present in the color-color
plot (Fig.~\ref{the_class_col6}). The behavior is opposite  to that
of the classes resulting from classifying spectra with continua,
which are well separated in the color-color plot (Fig.~\ref{the_class_col4}),
but overlap in the [Fe/H]-$T_{\rm eff}$  diagram (Fig.~\ref{fe_vs_teff4}).
Note also that classes 5 and 6 occupy the same
region of the color-color plot and overlap in the 
$\log({\rm g})$-$T_{\rm eff}$ plot, but they have different metallicities.
These are F stars that approximately split according 
to their membership to the thick disk and  the halo \citep[][]{2006ApJ...636..804A},
therefore,  the classification 
provides a quick-look tool to separate disk and halo stars.
Class~12 is dominated by DA WDs (see below and
the class template  in Fig.~\ref{the_class_temp6mas}),
however, it shows up as extremely low metallicity in Fig.~\ref{density6}.
Since the SSPP does not deal with WDs, these must have been confused 
with A-type stars and analyzed as such, finding that they are best matched
with no metals.

\begin{figure*}
\includegraphics[width=0.8\textwidth]{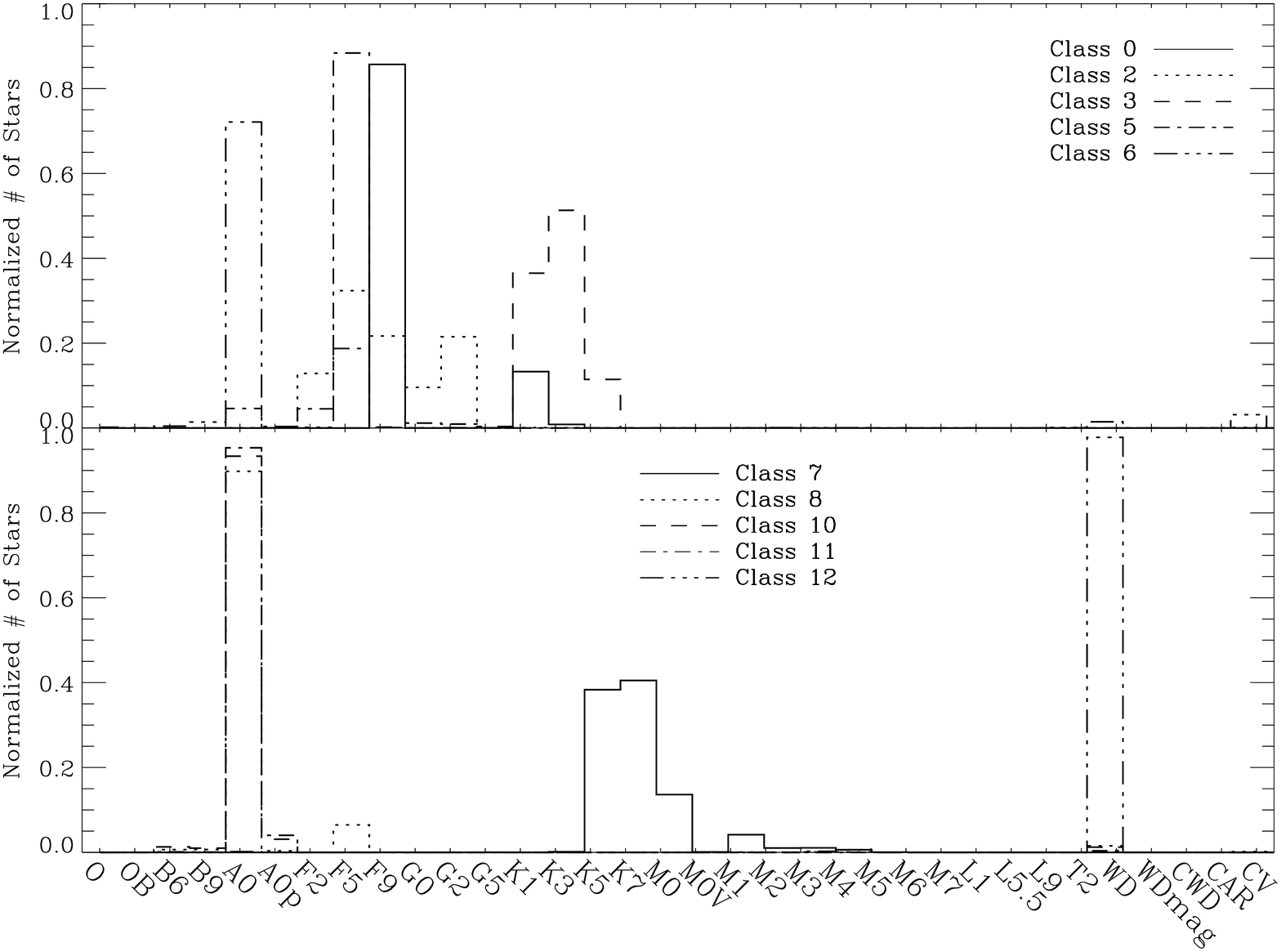}
\caption{
Histograms of the distribution of ELODIE MK types of selected k-means classes
corresponding to spectra without continuum (see Fig~\ref{the_class_temp6otro}). 
The histograms have been split in two panels to avoid overcrowding. 
Each one shows a set of k-means classes as described in the insets.
The histograms have been 
normalized to one, including objects where the 
MK type is not available. 
}
\label{mkclasses6}
\end{figure*}

\begin{figure*}
\includegraphics[width=0.81\textwidth]{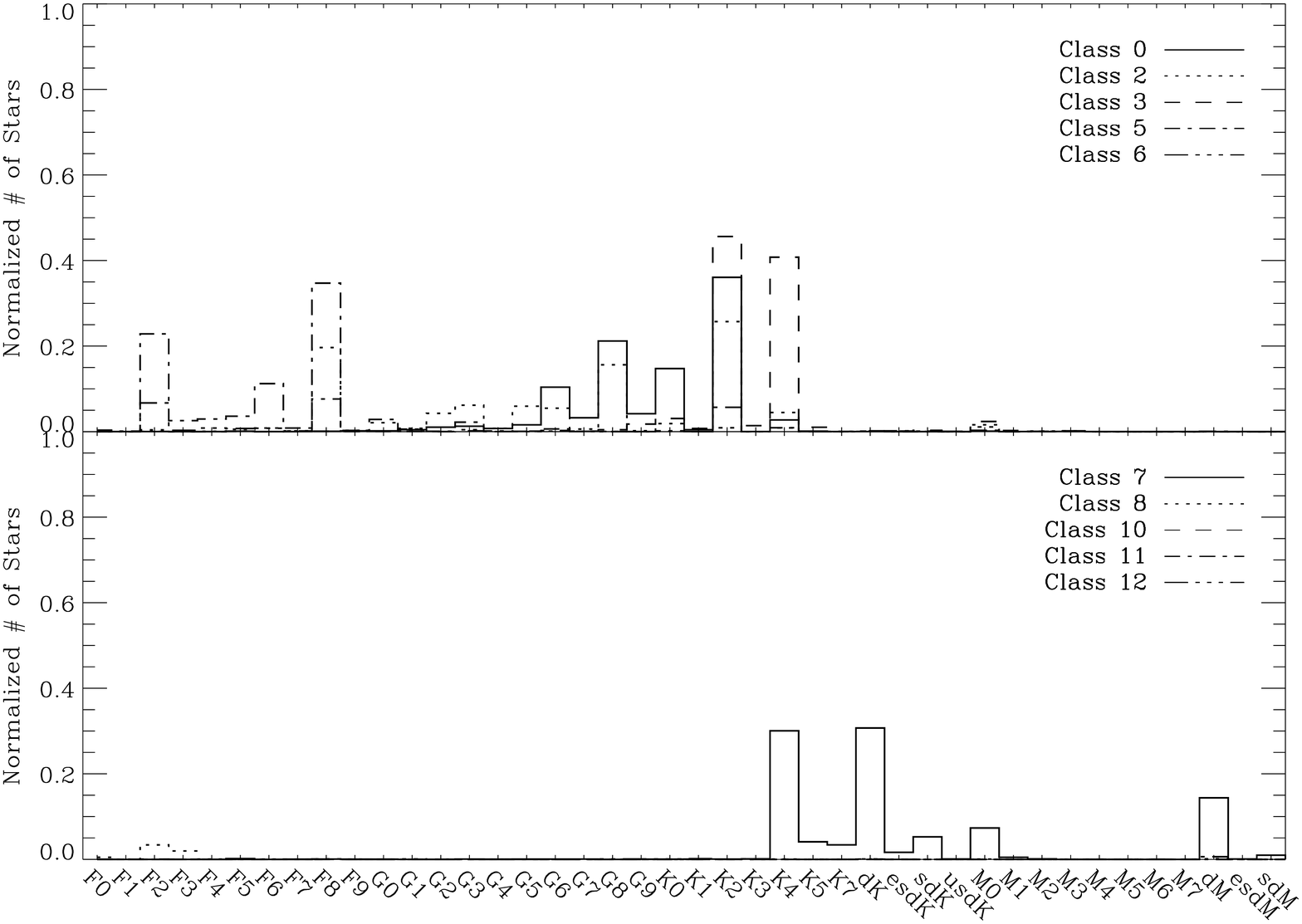}
\caption{
Similar to Fig.~\ref{mkclasses6} but showing Hammer MK types rather than 
ELODIE types. 
The histograms have been 
normalized to one, including objects where the 
type is not available. Classes 8, 10, 11 and 12 do not appear
because they contain hot stars with no associated
Hammer type. Note that class~0 collects types G and early K, whereas it appears
as formed by F9 stars in Fig.~\ref{mkclasses6}.
}
\label{mkclasses6_otro}
\end{figure*}

Figure~\ref{mkclasses6} shows the distribution
of MK spectral types corresponding to the k-means classes.
Note how each class tends to belong to a single spectral 
type, but not always.  Moreover, the correspondence seems to be
better than that for the classification including continuum
(cf., Fig.\ref{mkclasses4}). Note how class~12  is basically formed
by WDs, whereas classes 8, 10 and 11 are made of type A stars. 
The most numerous class~0 contains almost
exclusively F9 stars, 
as it also happens with the classification with continuum
(Fig.~\ref{mkclasses4}).
We think that the concentration of class~0 around a particular type is 
real, but the particular type  is not, since most stars selected by SEGUE are  
G-type rather than F-type (see Sect.~\ref{data}). There seems to be a problem 
with the MK typing based on ELODIE templates because, as expected, 
the Hammer MK types associated with class~0 are late G types. 
Figure~\ref{mkclasses6_otro} is equivalent
to Fig.~\ref{mkclasses6} but showing Hammer types 
(Sect.~\ref{data}). Class~0 corresponds to types between
G6 and K2. Note, in passing, that the classes of hot stars
have disappeared from the histograms since Hammer typing 
does not allocate classes to them.

Table~\ref{summary} gives a summary of the physical properties of the classes, namely,
it contains mean values and standard deviations for colors, temperatures, gravities
and metallicities.  (Class~13 has been excluded since it seems to be collecting faulty 
spectra.) Note that some classes present a fairly small range of physical parameters.
For instance, if a star is assigned to   class~0 then  we know its temperature, gravity and 
metallicity with standard deviations of 190~K, 0.25~dex, and 0.36~dex,  respectively.
These uncertainties are comparable to those associated
with other approaches currently used to estimate  effective 
temperature and gravity. 
This fact opens up the possibility of using k-means for a quick-look estimate
of the main physical parameters of the stars at a minimum computational
cost. Once the classes are known, assigning a new spectrum to 
one of them is virtually instantaneous -- it is just a matter of computing the 
difference between the new spectrum and the class templates,
 and then selecting the class of least rms deviation.


\begin{deluxetable}{cccccc}
\tablecolumns{6}
\tablewidth{0pc} 
\tablecaption{Properties of the stellar classes from spectra without continuum} 
\tablehead{ 
\colhead{Class}&\colhead{$g-r$}&\colhead{$u-g$}&\colhead{$T_{\rm eff}\tablenotemark{a}$}&$\log({\rm g})\tablenotemark{b}$
& [Fe/H]\tablenotemark{c}
}
\startdata
 0& 0.64$\pm$0.08&1.39$\pm$0.20&5370$\pm$190& 4.51$\pm$0.25&-0.53$\pm$0.36\\
 1& 0.53$\pm$0.10&1.03$\pm$0.17&5660$\pm$290& 4.13$\pm$0.47&-0.79$\pm$0.40\\
 2& 0.60$\pm$0.13&0.96$\pm$0.21&5450$\pm$340& 3.33$\pm$0.75&-1.71$\pm$0.53\\
 3& 0.87$\pm$0.13&1.86$\pm$0.29&4830$\pm$230& 4.61$\pm$0.22&-0.52$\pm$0.40\\
 4& 1.02$\pm$0.25&1.70$\pm$0.44&5260$\pm$440& 3.74$\pm$0.74&-0.55$\pm$0.50\\
 5& 0.36$\pm$0.12&0.74$\pm$0.13&6400$\pm$300& 3.98$\pm$0.35&-0.79$\pm$0.45\\
 6& 0.30$\pm$0.20&0.64$\pm$0.23&6500$\pm$330& 3.74$\pm$0.47&-1.86$\pm$0.61\\
 7& 1.31$\pm$0.17&2.24$\pm$0.33&4290$\pm$150& 4.39$\pm$0.34&-0.55$\pm$0.42\\
 8& 0.18$\pm$0.14&0.64$\pm$0.17&7250$\pm$340& 4.05$\pm$0.37&-0.90$\pm$0.62\\
 9& 0.69$\pm$0.16&1.31$\pm$0.35&5340$\pm$330& 4.24$\pm$0.47&-0.89$\pm$0.50\\
10& 0.02$\pm$0.14&0.59$\pm$0.17&8020$\pm$340& 4.12$\pm$0.40&-0.98$\pm$0.70\\
11&-0.14$\pm$0.08&0.49$\pm$0.12&8700$\pm$350& 3.80$\pm$0.47&-1.36$\pm$0.66\\
12&-0.27$\pm$0.13&0.35$\pm$0.18&8890$\pm$410& 4.67$\pm$0.17&-3.03$\pm$0.66
\enddata
\tablenotetext{a}{Effective temperature in K.}
\tablenotetext{b}{Gravity g in cm\,s$^{-2}$.}
\tablenotetext{c}{Metallicity in logarithmic scale referred to the Sun.}
\label{summary}
\end{deluxetable}

%
%
\section{Outliers}\label{outliers}
Having a classification automatically provides outliers, i.e.,
uncommon objects which therefore do not belong to 
any of the classes.
We can easily identify them since, in addition to assigning class 
memberships, our algorithm estimates the probability of 
belonging to the class  (see Sect.~\ref{algorithm}). 
Outliers are therefore those objects  whose
probability of belonging to their class is below a threshold.
We set the threshold to 0.01, which implies selecting as outliers
the 1\,\% spectra furthest  from their clusters centers.
The actual threshold is both arbitrary and unimportant, since 
our purpose was figuring out the type of spectra that do not
fit in  the classification. The adopted threshold renders some
2200 spectra, that were inspected individually. 

\begin{figure*}
\includegraphics[width=1.\textwidth,angle=0]{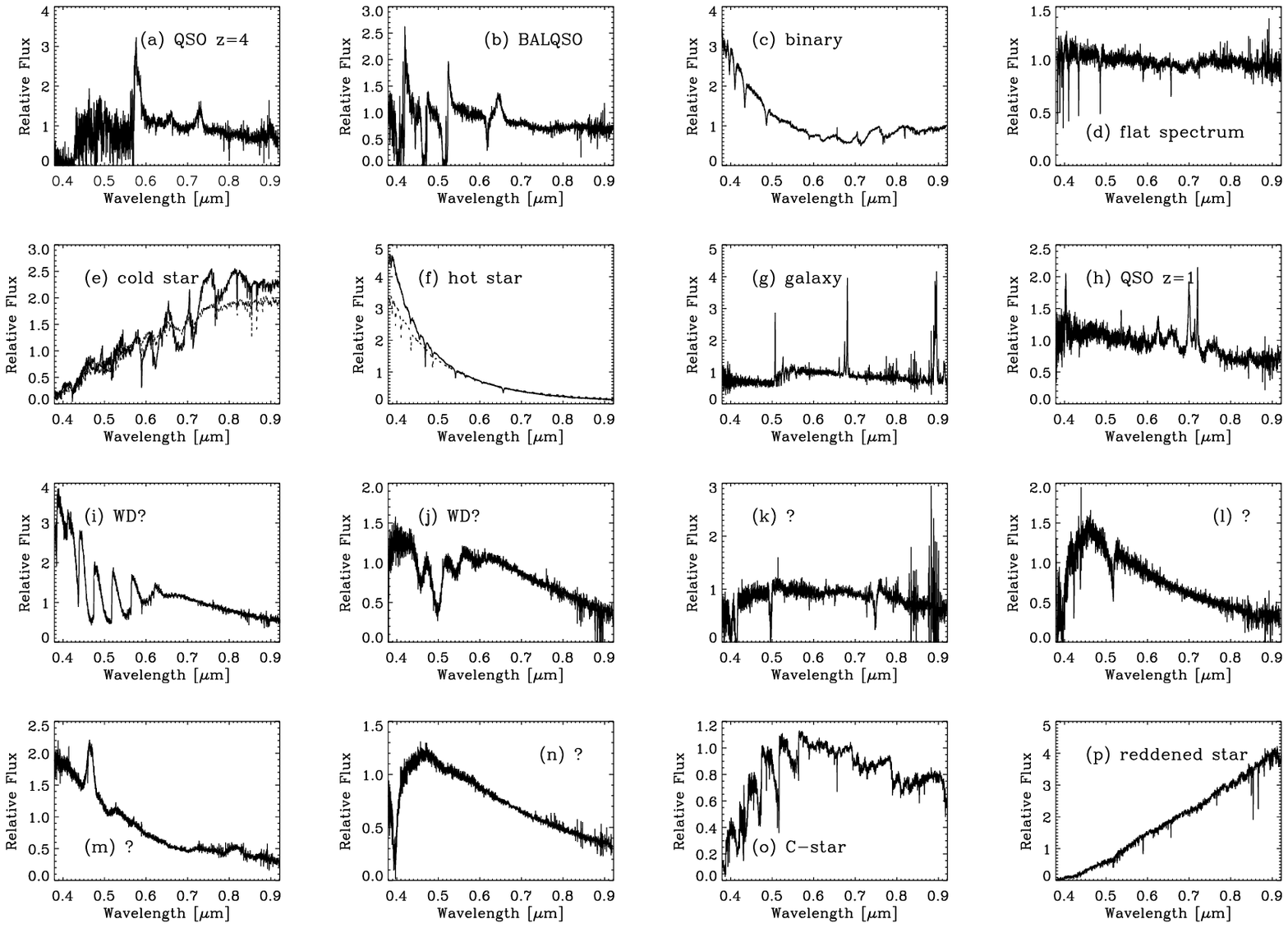}
\caption{
Examples of outliers of the classification that includes continuum.
(a) QSO at redshift around 4, so that  Ly$\alpha$ 
shows up at some 6000\,\AA .
(b) BAL~QSO, with broad absorption lines blue-shifted
with respect to their emission line counterparts. 
The largest signal corresponds to Ly$\alpha$.
(c) Composite spectrum formed by combining  light
from red and blue stars.
(d) Flat spectrum, showing signs of a hot star in the blue and
a red star in the red. They may be composite spectra like in (c) but
with the magnitudes of the two stars tuned to look spectrally flat.
(e) Spectrum of an extremely cold star -- the spectrum and 
the corresponding template are shown as solid and dotted lines,
respectively. 
(f) Star (the solid line) significantly hotter than its template (the dotted line).
(g) Star-forming galaxy at intermediate redshift.
(h) QSO at redshift around one.
(i--j) Carbon-rich WDs.
(k--n) Strange-looking spectra. 
They may be failures of the 
reduction pipeline, but they may be genuine abnormal 
objects as well. 
They include 
spectra with strong emission lines (m), 
or spectra with a single absorption line (n).
(o) Carbon stars.
(p) Dust-reddened blue stars.
}
\label{outlier1b}
\end{figure*}

We carried out
the  inspection for both classifications, the one
including continuum and the one without continuum.
In both cases the vast majority of the outliers are
noisy spectra or failures of the SDSS pipeline (e.g.,
gaps, unsuccessful removal of telluric lines,
mismatches between the red and the blue 
spectrograph arms, and so on). These problematic
spectra represent three-quarters of the outliers
for the classification using continuum, and a bit less when
continuum is removed. (This difference is to be 
expected since subtracting the continuum automatically
cancels many of the calibration problems.)
Here we focus on the outliers of the classification with 
continuum, although they are qualitatively similar to 
those of the classification with the continuum subtracted.
In this case  we count 548 genuine outliers. They  are
illustrated in Fig.~\ref{outlier1b}  and described 
in the following list ordered from more to less 
common:
\begin{itemize}
\item [-] Quasars (QSOs) at redshift between 2 and 4, so that Ly$\alpha$ appears
in the visible spectral range 
\citep[e.g., Fig.~\ref{outlier1b}a -- we  
identified the observed lines in the QSO spectrum 
template by][]{1991ApJ...373..465F}.
Most outliers are of this kind (some 320 or 60\% of the sample).
They represent only 0.3\% of QSOs in the 
latest SDSS QSO catalog \citep{2010AJ....139.2360S},
but they tend to appear in the redshifts where the SDSS 
identification algorithm has known problems (redshifts 2.9 and 3.2).
A bad redshift assignation at these particular redshifts
would explain the presence of a large number of QSOs 
contaminating  our stellar sample. 
%
\item [-] Broad Absorption Line (BAL) QSOs at the same high redshifts
(Fig.~\ref{outlier1b}b).
Those are believed to be AGNs with very rapid outflows 
(of a few  times 10$^4$\,km\,s$^{-1}$)
along the line of sight \citep{2009ApJ...692..758G}.
We get 33 of these objects, which represent 6.0\,\% of the outliers,
and 9.3\,\% of the QSO -- a percentage that seems to be normal
for BAL QSOs \citep[e.g.,][]{2009ApJ...692..758G}.
\item [-] Composite spectra of blue and red stars combined 
(e.g., Fig.~\ref{outlier1b}c, 
where H$\alpha$ shows up in emission).
They may be genuine
binary systems with the two stars gravitationally bounded, or just
two or more stars that happen to be along our line of sight. 
We ignore why these stars appear as outliers rather than elements of
existing classes (classes~20 or 21 in Fig.~\ref{the_class_temp4}), 
but it may be due to having an excess of blue upturn 
as compared to the template spectra.
\item [-] Flat spectra, showing absorption line features characteristics
of hot stars in the blue, 
and of cold stars in the red  (e.g., Fig.~\ref{outlier1b}d). 
They may be composite spectra like in Fig.~\ref{outlier1b}c,
but with the luminosities of the 
stars  fine-tuned so that the combination looks spectrally flat.
\item [-] Extreme spectra. They look-like the corresponding templates,
	but seem to be extreme cases (e.g., extreme colors or 
	particularly deep absorption lines).
Figure~\ref{outlier1b}e shows a particularly cold star or sub-stellar
object
 -- the spectrum and 
the corresponding template are shown as solid and dotted lines,
respectively. Figure~\ref{outlier1b}f shows the spectrum of a
star (the solid line) hotter than its template (the dotted line).
\item [-] Star-forming galaxies at intermediate redshifts (e.g., Fig.~\ref{outlier1b}g).
\item [-] QSOs at redshifts around one (e.g., Fig.~\ref{outlier1b}h).
\item [-] Strongly dust-reddened blue stars (e.g., Fig.~\ref{outlier1b}p).
\item[-]
Carbon rich WDs with strong C$_2$ bands (e.g., Fig.~\ref{outlier1b}i
and \ref{outlier1b}j) -- see the observed and 
synthetic  spectra in \citet{1984ApJ...284..257W}.
\item[-]
Carbon stars (e.g., Fig.~\ref{outlier1b}o), where the photospheric opacity is dominated by 
C-bearing molecules. The carbon, dredged up to the 
photosphere, comes from the He burning shell 
characteristic of low mass stars during their late stages of evolution
\citep[e.g.,][]{2009A&A...503..913A}.  
Figure~\ref{outlier1b}o should be compared with C-star spectra in,
e.g., \citet[][Fig.~4]{2001A&A...371.1065L} and  \citet[][Fig.~3]{2009A&A...503..913A}.
\item [-] Strange-looking spectra. They may be failures of the 
reduction pipeline, but they may be genuine abnormal 
objects as well. 
Figures~\ref{outlier1b}k--\ref{outlier1b}n show 
a few of them, chosen only when they are not the sole
representative of its class.
They include spectra with strong emission 
lines (e.g., Fig.~\ref{outlier1b}m), 
or spectra with a single absorption line
(e.g., Fig.~\ref{outlier1b}n).
\end{itemize}

%
\section{Discussion and Conclusions}\label{conclusions}
The traditional approach to classify stellar spectra has to be 
adapted to process the volume of data produced by
large surveys underway (see Sect.~\ref{intro}).  
There is a need for new automated techniques of analysis. 
In this work we explore the
use of the algorithm k-means for the task, i.e., 
as a tool for the  automated unsupervised classification of massive stellar
spectrum catalogs. The algorithm has already proven its 
potential to fast processing other astronomical spectra (Sect.~\ref{intro}), and 
we expected it to be useful in this context as well.

For this exploratory application of k-means, we selected the data set of stellar 
spectra associated with the SEGUE and SEGUE 2 programs. Even though
it is not a fair sample of the Milky Way stellar populations, it contains 
a rich variety of stellar types and so it is a good test-bench for the
classification algorithm. After discarding  faulty cases, 
the reference dataset consists of 173,390 
stellar spectra from 3800 to 9200\,\AA\  sampled in 3849 wavelengths.
Therefore, 
the problem for  k-means is  to find
clusters among 173,390 vectors 
defined in a 3849-dimensional space.  
The full data set occupies 6.1\,GB, and is classified using  a
standard up-to-date workstation in some two hours (see Sect.~\ref{repeatability}). 

We apply the classification to the original spectra and also 
to the spectra with the continuum removed. The latter data set 
contains only spectral lines, and it is less dependent on 
observational and instrumental problems like dust extinction, 
spectro-photometric mis-calibrations,
or  failures in the reduction pipeline.

The classification of the spectra with continuum renders
16 major classes, with 99\,\% of the objects, and ten minor 
classes with the remaining 1\,\% (Sect.~\ref{class_with}). Roughly speaking, the 
stars are split according to their colors, with enough 
finesse to distinguish dwarf and giant stars 
(Fig.~\ref{the_class_col4}).  Figure~\ref{the_class_temp4}
shows the template spectra representative of all the classes:
there are classes for
WDs (class~15),
A-type stars (class 9),
F-type stars (class~0),
K-type stars (class~8),
M-type stars (class~23),
dust-reddened intrinsically blue stars (class~18),
binary systems (class 21),
and even classes with faulty spectra (classes 17 and 19). 
 It must be stressed, however,
that there is not a one-to-one correspondence
between the classes we derived and the MK types.
Often our classes mix-up several MK types, and vice versa. 
The classification is able to separate stars with similar
temperatures but different surface gravities
(compare classes 0 and 3 in Fig.~\ref{density4}), but
has difficulties to separate stars with different
metallicities (Fig.~\ref{fe_vs_teff4}). 

The classification of spectra without continuum renders less classes
(Sect.~\ref{class_out}) -- 13 major classes 
and only 1 minor class that probably collects faults of the
reduction pipeline (Figs.~\ref{the_class_temp6} and 
\ref{the_class_temp6mas}).
In this case the color separation is not so sharp as it is
for the classification with continuum included 
(cf., Figs~\ref{the_class_col4} and \ref{the_class_col6}).
However, it is able to separate stars in classes with
the same effective temperatures but different metallicities (Fig.~\ref{fe_vs_teff6}).
The behavior is opposite  to that
of the classes resulting from classifying spectra with continua,
which are well separated in the color-color plot
but overlap in [Fe/H]~vs~$T_{\rm eff}$.

Some classes include starts with  a fairly small range of physical 
parameters, as assigned by the SSPP. 
The mean value and dispersion of the effective temperature,
surface gravity, and metallicity of the classes without continuum 
are listed in Table~\ref{summary}. A small dispersion
implies that our classification can be used to estimate
the main physical parameters of the stars at a minimum computational
cost. One only has to assign the problem spectrum to one of the existing
classes, e.g., to the one of minimum residual. Then the properties 
of the class can be passed on to the new spectrum, thus providing its
main physical properties. For example,
if the problem spectrum
happens to belong to class~0, then we know its temperature, gravity, and 
metallicity within a standard deviation of 190~K, 0.25~dex, and 0.36~dex,  
respectively. These uncertainties are probably upper
limits since the estimate of the physical parameters by the SSPP 
has  non-negligible internal errors
that are included in the dispersions.
Note that the uncertainties are comparable with those associated
with other approaches currently used to estimate  effective 
temperature and gravity, which are far more time consuming.
Moreover, since we derive physical parameters from spectra
without continuum, the estimates are fairly robust against 
dust reddening and other observational issues, which  are often 
a serious  problem  when dealing with stars at low galactic  latitudes.

The classification also provides a means of
finding rare but scientifically interesting
objects, e.g., unusually low metallicity stars, odd spectral types, etc.
By definition, rare objects must be outliers of any classification, 
otherwise they would be common and would have classes associated 
to them.
Our rendering of k-means gives the goodness of the assignation, i.e.,
the probability that each star belongs to the class it  has been 
assigned to. Therefore, the outliers of the classification are easy to pinpoint as those
spectra whose probability of a correct assignation is low 
enough. The nature of the outliers thus selected was
examined in Sect.~\ref{outliers} -- see Fig.~\ref{outlier1b}.   
Most outliers are faulty data or failures of the SDSS reduction 
pipeline.
The remaining 25\,\%  is firstly formed by
high redshift QSOs. Since they are in the appropriate redshift range, we
speculate that these QSO may be those lost by a known problem in the SDSS
QSO identification algorithm.  There is a large number of outliers
corresponding to composite spectra formed by either real or fake double 
or multiple stellar systems. The spectrum is that of a hot star in the blue and
a cold star in the red. There are also reddened stellar spectra, and galaxy spectra.
Finally, there are odd spectral types whose nature we did not manage to
figure out, and which we plan to observe in follow up work.    

One obvious use of the present classification is identifying
spectra having instrumental problems or being produced
by flaws of the reduction pipeline. We find classes containing
faulty spectra when the problem is common, and then we find 
faulty spectra as outliers of the classification when the problem
is unusual.

Stellar spectra are known to be highly compressible so that they can be 
characterized using only a few independent parameters (see Sect.~\ref{intro}). 
Then the fact that the classes present a regular behavior was somehow 
expected,  and this fact is not the main outcome of our exercise. 
Instead, our exploratory work shows k-means to provide a viable tool 
for the systematic classification of  large data sets of  stellar 
spectra. Moreover,  there
is plenty of room for improving the procedure, i.e., for
upgrades that have not been considered in the paper, but which
may be of interest in future uses.
One can focus the classification in a particular spectral
range (or set of ranges) particularly sensitive to the
physical parameter one wants to select
(say, the metallicity if searching for classes of extremely metal 
poor stars). 
Then the resulting classes would emphasize this particular aspect of the spectra.
Obviously, using smaller spectral ranges for classification also
speeds up the procedure.
One can also resort to nested
k-means classifications, where the spectra of a given class are
separated into subclasses. This can be used to fine-tuning separation.

Finally, we want to indicate that the template spectra from the classifications
with and without continuum are publicly available\footnote{{\tt 
ftp://stars:kmeans@ftp.iac.es}}.

%
%



%
%
\begin{acknowledgements}
Thanks are due to J.~Casares, L.~Girardi, and C.~Ramos-Almeida  
for help with deciphering 
the nature of some of the outliers,
and to  I.~Ferreras  and B.~Gustafsson 
for illuminating discussions.
In addition, we thank the referee, C. Bailer-Jones, for his constructive
criticisms that tempered the statements made in the paper.
This work has been partly funded by the Spanish Ministry for Science, 
project AYA~2010-21887-C04-04.          
%
JSA is member of the Consolider-Ingenio 2010 Program, grant 
MICINN CSD2006-00070: First Science with GTC.
Funding for SDSS, SDSS-II, and  SDSS-III has been provided by the Alfred P. Sloan Foundation, 
the Participating Institutions, the National Science Foundation, 
and the U.S. Department of Energy Office of Science.
SDSS-III is managed by the Astrophysical Research Consortium for the Participating Institutions 
of the SDSS-III Collaboration including the University of Arizona, the Brazilian Participation Group,
 Brookhaven National Laboratory, University of Cambridge, Carnegie Mellon University, University of Florida, 
the French Participation Group, the German Participation Group, Harvard University,
 the Instituto de Astrofisica de Canarias, the Michigan State/Notre Dame/JINA Participation Group, 
Johns Hopkins University, Lawrence Berkeley National Laboratory, Max Planck Institute for Astrophysics, 
Max Planck Institute for Extraterrestrial Physics, New Mexico State University, 
New York University, Ohio State University, Pennsylvania State University, University of Portsmouth,
 Princeton University, the Spanish Participation Group, University of Tokyo, 
University of Utah, Vanderbilt University, University of Virginia, 
University of Washington, and Yale University.
{\it Facilities:} \facility{Sloan (DR8, spectra)
}
\end{acknowledgements}

%
%

\end{document}